\documentclass{ieeeaccess}
\makeatletter
\renewcommand{\headerlogo}{}
\renewcommand{\headerlogoall}{}
\makeatother
\usepackage{cite}
\usepackage{amsmath,amssymb,amsfonts}
\usepackage{algorithmic}
\usepackage{graphicx}

\usepackage{xfrac}
\usepackage{textcomp}
\usepackage{xcolor}
\usepackage{array}
\usepackage{booktabs}
\usepackage{float}
\usepackage{tabularx}
\usepackage{cuted}
\usepackage{capt-of}
\usepackage{dblfloatfix}
\usepackage{placeins}
\usepackage{algorithm}
\usepackage{algorithmic}
\usepackage{eurosym}

\definecolor{workflowblue}{RGB}{18,38,83}
\definecolor{workflowgreen}{RGB}{214,239,205}
\definecolor{workflowlight}{RGB}{242,247,250}
\definecolor{workflowgray}{RGB}{248,248,248}

\definecolor{RED}{RGB}{255,0,0}
\definecolor{BLUE}{RGB}{0,0,255}
\definecolor{GREEN}{RGB}{0,128,0}

\def\BibTeX{{\rm B\kern-.05em{\sc i\kern-.025em b}\kern-.08em
    T\kern-.1667em\lower.7ex\hbox{E}\kern-.125emX}}

\makeatletter
\providecommand{\xfigwd}{\columnwidth}
\makeatother

\begin{document}
\history{Date of publication xxxx 00, 0000, date of current version xxxx 00, 0000.}
\doi{}

\title{Towards Scalable Quantum Unit Commitment in Power Systems: Encoding and Complexity Reduction Techniques}

\author{\uppercase{Marc Carrillo-Muñoz}\authorrefmark{1},
\uppercase{Sergio Martínez-Hermida}\authorrefmark{1}, 
\uppercase{Sara Barja-Martínez}\authorrefmark{1},
\uppercase{Antonio E. Saldaña González}\authorrefmark{1}, and
\uppercase{Mònica Aragüés-Peñalba}\authorrefmark{1,2}}

\address[1]{Dpt. Enginyeria Elèctrica, CITCEA-UPC, Universitat Politècnica de Catalunya, Barcelona, Spain}
\address[2]{Senior Member IEEE}

\tfootnote{The authors thankfully acknowledge the computer resources at Quantum-Blue and Quantum-Red and the technical support provided by Barcelona Supercomputing Center (BSC) and Finisterrae-QMIO-Qulacs, supported by the Centro de Supercomputación de Galicia (CESGA) under activity IM-2025-3-0029 "Quantum and hybrid quantum-classical algorithms for power system applications". This work was part of the I+D+i project OPERA (Operation and Planning tools for Enabling Renewable distribution systems Acceleration based on emerging technologies for sustainable computing) with reference PID2024-160822OB-I00 funded by the Ministerio de Ciencia, Innovación y Universidades from Spain. The work from Mònica Aragüés was supported by the ICREA Academia Program.}

\markboth
{Author \headeretal: \textit{Marc Carrillo i Muñoz}}
{Author \headeretal: \textit{Towards Scalable Quantum Unit Commitment: Encoding and Complexity Reduction Techniques}}

\corresp{Corresponding author: Marc Carrillo-Muñoz (email: marc.carrillo.munoz@upc.edu).}

\begin{abstract}
This paper develops a Quantum optimisation Unit Commitment (UC) framework that benefits from Pauli Correlation Encoding (PCE) as a qubit reduction preprocessing technique and Sample-based Quantum Diagonalisation (SQD) as a hybrid post-processing refinement method. The UC problem is formulated in binary form, then transformed into a QUBO problem and mapped onto an Ising representation suitable for adiabatic quantum evolution within the Qibo framework. This study analyses the how PCE and SQD techniques affect qubit scalability, computational effort, feasibility, and probability of recovering the optimal solution as the number of generators increases. The results show that PCE reduces the effective size of the encoded problem, SQD improves the concentration of probability on feasible and optimal solutions, and their combined application provides the best overall behaviour in several scalable case studies, achieving more reliable solutions with lower effective computational cost than the unreduced formulation. The repository is available open source in Github (link is not yet available until work is published. However, if requested by the reviewers, we can make it open for a scific time window, if requested)
\end{abstract}


\titlepgskip=-15pt
\maketitle

\newcommand{\notesection}[1]{%
\vspace{0.5mm}
\noindent\textbf{#1}\par
\vspace{0.7mm}
\hrule
\vspace{0.7mm}
}

\newcommand{\noteitem}[2]{%
\noindent
\makebox[0.25\columnwidth][l]{#1}%
\parbox[t]{0.76\columnwidth}{#2}\par
\vspace{0.25mm}
}

\begin{scriptsize}
\setlength{\parindent}{0pt}

\noindent\textbf{Main notation used in the proposed quantum UC workflow.}
\vspace{1mm}

\notesection{Sets, indices and UC data}
\noteitem{$\mathcal{G}$, $g$}{Set and index of generating units}
\noteitem{$\mathcal{T}$, $t$}{Set and index of UC time periods}
\noteitem{$\mathcal{D}$, $i$}{Set and index of demand/load contributions}
\noteitem{$N$}{Number of original logical variables/qubits}
\noteitem{$P_g$ [MW]}{Available power of generator $g$ in the commitment based model}
\noteitem{$P_g^{\min},P_g^{\max}$ [MW]}{Minimum and maximum generation limits}
\noteitem{$D_t$, $D$ [MW]}{Demand at time $t$ and simplified single period demand}
\noteitem{$d_{i,t}$ [MW]}{Demand contribution of load $i$ at time $t$}
\noteitem{$c_g$}{Generation cost coefficient of unit $g$}
\noteitem{$C_g^{\mathrm{gen}}(\cdot)$}{Generation cost function}
\noteitem{$C_g^{\mathrm{su}}$}{Start-up cost coefficient}

\notesection{UC, QUBO and Ising formulation}
\noteitem{$u_{g,t}$, $u_g$}{Binary commitment status of generator $g$}
\noteitem{$y_{g,t}$}{Binary start-up indicator of generator $g$ at time $t$}
\noteitem{$p_{g,t}$ [MW]}{Dispatch power of generator $g$ at time $t$}
\noteitem{$\mathbf{x}\in\{0,1\}^{N}$}{Binary optimisation vector}
\noteitem{$x_i$}{Binary variable associated with logical qubit $i$}
\noteitem{$z_i,z_g\in\{-1,+1\}$}{Classical spin variable associated with a binary commitment variable}
\noteitem{$Z_i,Z_g$}{Pauli-\(Z\) operator associated with a logical spin}
\noteitem{$A$}{Raw QUBO penalty factor}
\noteitem{$Q(\mathbf{x})$}{QUBO objective function}
\noteitem{$\alpha^C$}{Cost normalisation factor}
\noteitem{$\alpha^P$ [MW]}{Power-balance normalisation factor}
\noteitem{$\tilde c_g,\tilde P_g,\tilde D_t$}{Normalised cost, power and demand}
\noteitem{$\lambda_{\mathrm{auto}}$}{Automatic penalty baseline}
\noteitem{$\lambda$, $\eta$}{Penalty coefficient and dimensionless multiplier}
\noteitem{$\lambda_{\min}$}{Minimum penalty required to preserve the optimal feasible QUBO ground state}
\noteitem{$h_i$, $J_{ij}$}{Ising local fields and pairwise couplings}
\noteitem{$E_0$}{Constant energy shift in the Ising Hamiltonian}
\noteitem{$H_1\equiv H_{\mathrm{Ising}}$}{Original UC problem Hamiltonian}

\notesection{PCE parameters and variables}
\noteitem{$m$}{Number of reduced/effective qubits, $m<N$}
\noteitem{$\mathbf{s}\in\{-1,+1\}^{m}$}{Reduced spin assignment}
\noteitem{$s_q$}{Reduced spin variable associated with reduced qubit $q$}
\noteitem{$k$}{Allowed PCE correlation orders}
\noteitem{$k_{\max}$}{Maximum correlation order allowed by \texttt{k}}
\noteitem{$S_i$}{Reduced-qubit support set encoding logical operator $Z_i$}
\noteitem{$|S_i|$}{Support size of the Pauli string encoding $Z_i$}
\noteitem{$\Pi_i$}{Pauli string replacing logical operator $Z_i$}
\noteitem{$f_{\mathrm{PCE}}(\cdot)$}{Map from reduced spin assignments to logical spin assignments}
\noteitem{$\hat z_i$}{Decoded logical spin after PCE}
\noteitem{$H_{\mathrm{PCE}}$}{PCE-compressed Hamiltonian}
\noteitem{$C_{\mathrm{feas}}$}{Feasible-state coverage of the selected representable or retained subspace}
\noteitem{$w(\mathbf{u})$}{Weight assigned to feasible UC configuration \(\mathbf{u}\) during PCE encoding selection}
\noteitem{$\lambda_{\mathrm{cost}}$}{PCE cost-prioritisation weight used during encoding selection}

\notesection{Adiabatic evolution and sampling}
\noteitem{$n_{\mathrm{eff}}$}{Effective number of qubits used in the quantum evolution}
\noteitem{$H_0$}{Transverse-field driver Hamiltonian}
\noteitem{$X_q$}{Pauli-\(X\) operator acting on qubit $q$}
\noteitem{$H_{\mathrm{problem}}$}{Hamiltonian evolved by the quantum solver}
\noteitem{$H(t)$}{Time-dependent adiabatic Hamiltonian}
\noteitem{$T$}{Final adiabatic evolution time}
\noteitem{$\Delta t$}{Simulation time step}
\noteitem{$\tau$}{Adiabatic schedule/evolution scaling parameter}
\noteitem{$|\psi\rangle^0$}{Initial equal-superposition state}
\noteitem{$|\psi\rangle^{\mathrm{opt}}$}{Target optimal/low-energy final state}
\noteitem{$b$}{Computational-basis bitstring}
\noteitem{$f(b)$}{Measurement frequency of bitstring $b$}
\noteitem{$\Omega$}{Set of distinct observed bitstrings}
\noteitem{$d=2^{n_{\mathrm{eff}}}$}{Effective Hilbert-space dimension}
\noteitem{$U=|\Omega|$}{Number of distinct measured bitstrings}

\notesection{SQD post-processing}
\noteitem{$\rho_{\mathrm{SQD}}$}{Target retained fraction for SQD subspace size}
\noteitem{$K_{\mathrm{target}}$}{Target SQD subspace size}
\noteitem{$\mathcal{S}_{\mathrm{SQD}}$, $S$}{Sampled subspace retained by SQD}
\noteitem{$M$, $|\mathcal{S}_{\mathrm{SQD}}|$}{Number of bitstrings retained by SQD}
\noteitem{$\rho_{\mathrm{eff}}$}{Realised retained fraction, \(M/d\)}
\noteitem{$r_{\mathrm{SQD}}$}{Realised SQD dimensional reduction}
\noteitem{$\beta$}{SQD auxiliary-Hamiltonian mixing parameter}
\noteitem{$H^{*}(\beta)$}{SQD auxiliary Hamiltonian}
\noteitem{$H_{\mathrm{sub}}$}{SQD projected Hamiltonian in the sampled subspace}
\noteitem{$v_0$, $e_0$}{Ground eigenvector and eigenvalue of $H_{\mathrm{sub}}$}
\noteitem{$v_{0,j}$}{Amplitude of sampled state $b^{(j)}$ in the SQD ground eigenvector}
\noteitem{$|\Psi\rangle^{\mathrm{SQD}}$}{SQD-refined state in the sampled subspace}

\notesection{Decoding and evaluation}
\noteitem{$\mathcal{D}(\cdot)$}{Decoding map to the original UC space}
\noteitem{$\hat{\mathbf{x}}$}{Decoded binary UC schedule}
\noteitem{$\hat u_g$, $\hat u_{g,t}$}{Decoded binary generator commitment decision}
\noteitem{$\hat{\mathbf{u}}$}{Decoded UC commitment schedule}
\noteitem{$\hat{\mathbf{u}}^{\star}$}{Best feasible decoded schedule}
\noteitem{$P_{\mathrm{eff}}(b)$}{Effective probability before decoding}
\noteitem{$P_{\mathrm{UC}}(\hat{\mathbf{x}})$}{Aggregated probability of decoded UC schedule}
\noteitem{$\mathcal{F}$}{Set of feasible decoded schedules}
\noteitem{$\mathcal{O}$}{Set of optimal or best-known decoded schedules}
\noteitem{$r(\hat{\mathbf{u}})$ [MW]}{Demand-balance residual of decoded schedule}
\noteitem{$C(\hat{\mathbf{u}})$}{Cost of decoded UC schedule}
\noteitem{$C^{\star}$, $C_{\mathrm{best}}$}{Reference optimal cost and best recovered cost}
\noteitem{$P_{\mathrm{feas}}$}{Probability mass assigned to feasible schedules}
\noteitem{$P_{\mathrm{opt}}$}{Probability mass assigned to optimal or best-known schedules}
\noteitem{$R_q$ [\%]}{Relative qubit reduction}
\noteitem{$\mathrm{gap}$ [\%]}{Relative optimality gap}

\notesection{Reported computational metrics}
\noteitem{$T_{\mathrm{tot}}$ [s]}{Total measured runtime of the selected workflow}
\noteitem{$T_{\mathrm{PCE}}$ [s]}{PCE preprocessing or compressed-model construction time}
\noteitem{$T_{\mathrm{AE}}$ [s]}{Adiabatic-evolution execution time}
\noteitem{$T_{\mathrm{SQD}}$ [s]}{SQD post-processing time}
\noteitem{$P_{\mathrm{feas}}^{\mathrm{AE}}$}{Feasible probability after direct adiabatic evolution}
\noteitem{$P_{\mathrm{opt}}^{\mathrm{AE}}$}{Optimal/best probability after direct adiabatic evolution}
\noteitem{$P_{\mathrm{feas}}^{\mathrm{SQD}}$}{Feasible probability after SQD post-processing}
\noteitem{$P_{\mathrm{opt}}^{\mathrm{SQD}}$}{Optimal/best probability after SQD post-processing}
\noteitem{$P_{\mathrm{feas}}^{\mathrm{PCE}}$}{Feasible probability after PCE-based adiabatic evolution}
\noteitem{$P_{\mathrm{opt}}^{\mathrm{PCE}}$}{Optimal/best probability after PCE-based adiabatic evolution}
\noteitem{$P_{\mathrm{feas}}^{\mathrm{PCE+SQD}}$}{Feasible probability after PCE and SQD}
\noteitem{$P_{\mathrm{opt}}^{\mathrm{PCE+SQD}}$}{Optimal/best probability after PCE and SQD}

\end{scriptsize}

\section{Introduction}

Quantum computing (QC) is increasingly being explored as an engineering paradigm for encoding and solving computationally demanding optimisation problems. A central challenge is to design quantum workflows that map real applications into suitable Hamiltonians, reduce the required number of qubits, and recover valid solutions after measurement and decoding. Power systems operation provides a relevant benchmark domain because problems such as AC Optimal Power Flow (AC-OPF), Optimal Transmission Switching (OTS), stochastic or security-constrained Unit Commitment (UC), and large scale planning combine physical constraints, discrete decisions, and rapidly growing problem dimensions. Among them, UC is especially suitable for quantum optimisation because its binary commitment structure can be naturally mapped to QUBO and Ising formulations, while its feasibility constraints make solution recovery non-trivial. Recent studies have reviewed the opportunities of QC in power systems, while also highlighting the hardware, algorithmic, and scalability limitations that still restrict its practical deployment \cite{ganeshamurthy2024next, Ullah2022}. The growing availability of quantum simulators, software frameworks, and cloud-accessible platforms is now enabling more systematic evaluation of end-to-end quantum optimisation pipelines.

Within power systems, optimisation has become one of the main areas of interest for quantum methods \cite{Liu2023}. DC Optimal Power Flow (DC-OPF), power flow models, and related economic dispatch problems have been used as early test cases because they can be reformulated in binary or quadratic form \cite{amani2023quantum, Carrillo, liu2024quantum}. However, these problems are often treated as proof of concept benchmarks and do not always reflect the full combinatorial structure of operational planning. In parallel, peer-to-peer (P2P) energy trading and energy communities have also attracted attention because of their market-based and decentralised decision layers \cite{bucher2025grid, o2023quantum}. These studies are relevant, but they are usually centred on transaction allocation or local coordination rather than on the qubit-scaling limitations of large constrained objective functions. UC, on the other hand, provides a more challenging benchmark as a combinatorial problem; the on/off scheduling of generating units must satisfy demand and operational constraints while minimising operating cost \cite{nikmehr2022quantum, koretsky2021adapting}. Its binary and constrained structure makes it a useful test case for analysing whether quantum optimisation methods can recover feasible and high-quality solutions as the problem size increases.

A key limitation of current quantum optimisation is the restricted number of available qubits, together with the cost of simulating or executing large Hamiltonians. This issue is particularly relevant in power systems problems, where the number of binary variables can grow quickly with the number of assets, time periods, and constraints. To address this limitation, recent works have proposed qubit-efficient and hybrid strategies. In UC, for instance, decomposition and qubit-efficient formulations have been studied to reduce the size of the problem or adapt it to quantum annealing resources \cite{ling2025hybrid, hong2025qubit}. These approaches are valuable, but they do not directly investigate a gate-based Hamiltonian-compression strategy in which the original logical variables of the UC problem are encoded into a smaller correlated-qubit representation before the Adiabatic Quantum Evolution (AQE), that evaluates the evolution of the states through the Hamiltonian.

More broadly, qubit-reduction methods such as Pauli Correlation Encoding (PCE) have shown promise in constrained optimisation and combinatorial problems by representing logical variables in a reduced Pauli-correlation space \cite{soloviev2025large, padin2026pauli}. However, these studies are not focused on UC, and they do not examine how the compressed Hamiltonian affects the recovery of feasible UC schedules after the quantum evolution. In parallel, Sample-based Quantum Diagonalisation (SQD) has been proposed as a hybrid quantum-classical post-processing method that refines solutions from a sampled subspace without requiring a new quantum evolution \cite{wang2025sample}. Still, SQD has not been systematically combined with a PCE-reduced UC Hamiltonian to assess whether post-processing can recover useful probability mass or compensate for feasibility losses introduced by aggressive qubit reduction.

The gap addressed in this paper is therefore the absence of an integrated, UC-oriented framework that jointly analyses qubit reduction, quantum evolution, post-processing, decoding, and evaluation in the original problem space. Existing quantum UC studies mainly focus on QUBO reformulations, decomposition strategies, or annealing-based implementations. At the same time, PCE + SQD have mostly been studied outside the UC context or as independent techniques. As a result, it remains unclear whether a PCE-compressed Hamiltonian can preserve the feasible and near-optimal structure of the original UC problem, and whether SQD can further improve the quality of the decoded solutions obtained from the reduced quantum dynamics. This distinction is important because reducing the number of qubits is not sufficient on its own: the reduced model must also retain the solution structure that matters for the original UC problem.

In power systems, quantum optimisation studies have mainly followed three directions. First, QUBO problem is defined solved through Ising formulation method, and QAOA-based formulations have been explored for UC, economic dispatch, OPF-related problems, and decentralised energy applications, providing early demonstrations of how operational constraints can be embedded into quantum-compatible objective functions \cite{amani2023quantum,Carrillo,liu2024quantum,nikmehr2022quantum,koretsky2021adapting,o2023quantum}. Second, qubit-efficient and decomposition-based approaches have been proposed to reduce the size of large scheduling problems and adapt them to current quantum-annealing or hybrid quantum-classical resources \cite{tan2021qubit_efficient_encoding,fuller2024quantum,kondo2025recursive,ling2025hybrid,muller2026quantum_annealing_unit_scheduling}. Third, recent encoding and post-processing techniques such as PCE and SQD aim to reduce the effective Hamiltonian dimension or refine the sampled solution space after quantum evolution \cite{sciorilli2025towards,soloviev2025large,padin2026pauli,robledo2025sqd,kaliakin2025implicit_sqd,shajan2025sqd_dmet}. However, these techniques have not yet been jointly analysed as a complete UC-oriented workflow in which qubit reduction, adiabatic evolution, SQD refinement, decoding, and original-space evaluation are studied together.

This paper studies how PCE together with SQD can be used as encoding and post-processing tools to reduce the effective complexity of quantum optimisation while preserving the ability to recover feasible UC solutions. The analysis focuses on scalable UC instances, where the number of generators, and therefore the number of qubits, increases progressively. The main contributions are:

\begin{itemize}

\item To formulate the UC problem as a binary optimisation model and map it to a QUBO/Ising Hamiltonian tailored for adiabatic quantum evolution within the Qibo \cite{qibo} framework, enabling its direct use in a quantum optimisation pipeline.

\item To analyse PCE as a qubit-reduction preprocessing strategy for solving the UC problem with AQC tehniques, assessing whether the compressed Hamiltonians preserve feasible and near-optimal schedules of the original problem.

\item To incorporate SQD as a hybrid post-processing step after the quantum evolution, evaluating its effect on feasibility, optimality, and the final probability assigned to relevant decoded UC solutions.

\item To demonstrate that the combined PCE+SQD workflow can improve the robustness of quantum UC solutions while requiring fewer effective qubits than the unreduced formulation. To the best knowledge of the authors, this is the first framework that combines PCE preprocessing and SQD post-processing for quantum optimisation in power systems UC.

\end{itemize}


\section{Methodology}
\label{sec:methodology}

\begin{figure}
    \centering
    \includegraphics[width=1\linewidth]{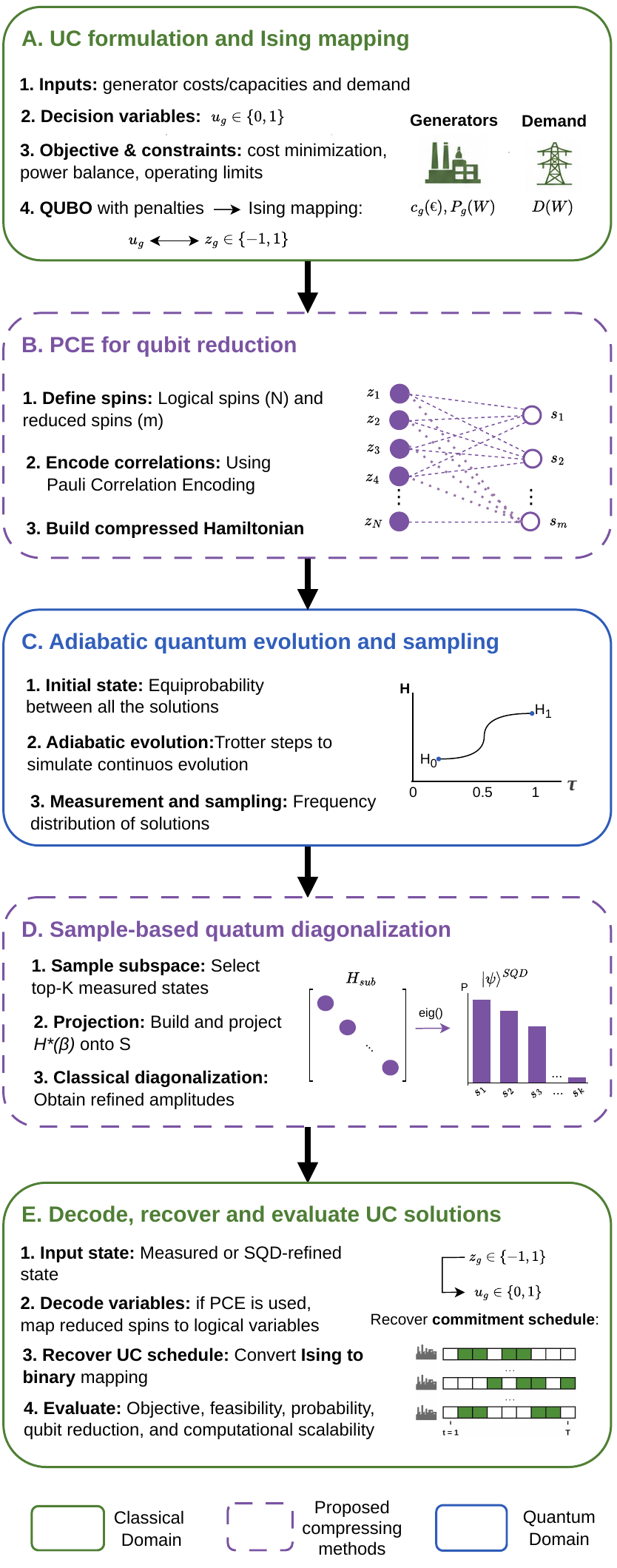}
    \caption{Methodology flowchart of the proposed quantum UC workflow.}
    \label{fig:methodology_workflow}
\end{figure}

This section describes the proposed quantum optimisation procedure for the UC problem. The proposed methodology is detailed in Fig.~\ref{fig:methodology_workflow}. The figure separates the stages that are always required from those that are not necessary to run the AQE but mandatory in order to reduce the complexity of the problem, which makes it easier to compare the baseline adiabatic approach with the PCE-only, SQD-only, and combined PCE + SQD configurations used in the case-study section.

The methodology is organised as an end-to-end quantum UC pipeline that starts from the classical UC formulation and finishes with the evaluation of decoded schedules in the original problem space. The first stage, common to all configurations, formulates the UC problem in binary form, maps it into a QUBO model, and obtains the corresponding Ising Hamiltonian.

From this common formulation, the workflow allows different configurations to be analysed under the same evaluation criteria. If PCE is activated, the original Hamiltonian is compressed before the quantum evolution and the solver acts on a reduced \(m\)-qubit representation. If PCE is not used, the original \(N\)-qubit Hamiltonian is evolved directly. After the adiabatic evolution and sampling stage, SQD can optionally be applied as a post-processing method to refine the probability distribution over the sampled states. Finally, all measured or refined states are decoded into UC schedules and evaluated using the original objective and feasibility constraints.

This structure allows four configurations to be compared consistently: the direct adiabatic baseline, PCE-only, SQD-only, and the combined PCE+SQD workflow. The rest of this section follows the order of Fig.~\ref{fig:methodology_workflow}. Subsection~\ref{subsec:uc_formulation} presents the UC formulation and Ising mapping. Subsection~\ref{subsec:PCE_methodology} describes the PCE-based qubit reduction. Subsection~\ref{subsec:adiabatic_solution} introduces the adiabatic evolution and sampling stage. Subsection~\ref{subsec:sqd_methodology} presents the SQD post-processing method. Subsection~\ref{subsec:decoding_uc} explains the decoding of quantum states into UC schedules. Last section also defines the evaluation criteria in the original UC space.

\subsection{Unit Commitment formulation and Ising mapping}
\label{subsec:uc_formulation}

Unit Commitment (UC) is a fundamental operational short-term planning problem in power systems. Its objective is to decide which generators should be turned on or off over a given time horizon in order to supply demand at minimum operating cost while satisfying the relevant operational constraints.

\begin{figure}[ht!]
    \centering
    \includegraphics[width=0.8\linewidth]{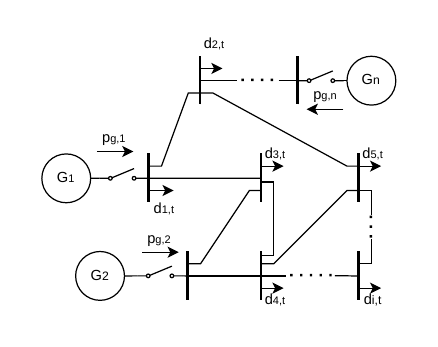}
    \caption{Generic electrical representation of the Unit Commitment problem.}
    \label{fig:uc_generic_representation}
\end{figure}

Figure~\ref{fig:uc_generic_representation} provides a representation of a generic electrical grid, its generators, delivering $p_{g,t}$ its loads consuming $d_{i,t}$ and their interconnectors. Generators are grid connected when $u_{g,t}$ is 1 and disconnected when $u_{g,t}$ is 0.


The general UC objective minimises operating costs over the time horizon, including production and start-up costs.

\begin{equation}
\min_{u,p}\;\; 
\sum_{t\in\mathcal{T}} 
\sum_{g\in\mathcal{G}} 
\Big( 
C^{\mathrm{gen}}_{g}(p_{g,t}) 
+ 
C^{\mathrm{su}}_{g}\,y_{g,t} 
\Big),
\end{equation}

$C^{\mathrm{gen}}_{g}$ is the generation cost function, $C^{\mathrm{su}}_{g}$ is the start-up cost coefficient, and $y_{g,t}\in\{0,1\}$ is a binary start-up indicator. The demand-balance constraint is imposed at every time step as

\begin{equation}
\sum_{g\in\mathcal{G}} p_{g,t} 
=
\sum_{i\in\mathcal{d}} d_{i,t}
=
D_t,
\qquad \forall t\in\mathcal{T},
\end{equation}

where $D_t$ is the total demand at time $t$. Commitment and dispatch are coupled through minimum and maximum generation limits,

\begin{equation}
P^{\min}_g\,u_{g,t} 
\le 
p_{g,t} 
\le 
P^{\max}_g\,u_{g,t},
\qquad \forall g\in\mathcal{G},\; t\in\mathcal{T}.
\end{equation}

The classical UC formulation introduces binary commitment variables

\begin{equation}
u_{g,t} \in \{0,1\}, \qquad \forall g\in\mathcal{G},\; t\in\mathcal{T},
\end{equation}

where $u_{g,t}=1$ indicates that generator $g$ is committed at time $t$, and $u_{g,t}=0$ otherwise. In addition, continuous dispatch variables

\begin{equation}
p_{g,t} \ge 0, \qquad \forall g\in\mathcal{G},\; t\in\mathcal{T},
\end{equation}

represent the generated power of each unit when active.

These constraints ensure that a generator can only produce power when it is committed and that its output remains within its operating range.

In this work, the UC problem is cast into a binary optimisation form suitable for quantum methods. Continuous dispatch decisions can be represented either by discretising $p_{g,t}$ into binary variables or by considering a simplified commitment-based UC formulation in which the main decision variable is the commitment vector $u_{g,t}$. In the latter case, each committed unit contributes an available power $P_g$, and the demand-balance condition is expressed directly in terms of the binary commitment variables. Furthermore, as ramp costs are not taken into account, the problem is solved independently at each time step, so the index $t$ is no longer required. The UC problem then becomes

\begin{equation}\label{uc_formulation}
\begin{aligned}
\min_{\mathbf{u}_g} \quad
& \sum_{g\in\mathcal{G}} c_g u_g \\
s.t. \quad
& \sum_{g\in\mathcal{G}} P_g u_g = D.
\end{aligned}
\end{equation}

The resulting binary model is transformed into a Quadratic Unconstrained Binary Optimisation (QUBO) problem by incorporating the demand-balance constraint as a penalty term, 
\begin{equation} 
\min_{\mathbf{u_{g}}} \left[ \sum_{g\in\mathcal{G}} c_g u_{g} 
+ A
\left( 
\sum_{g\in\mathcal{G}} P_g u_{g} - D 
\right)^2
\right]
\end{equation} 
where $A$ is a factor that scales the penalty of the non-feasible solutions over the optimisation function. This parameter is dependent on the problem, and its value is usually calculated with heuristic techniques. To make the formulation less sensitive to the numerical scale of the input costs and power-demand data, a normalised penalty model is used instead. The normalisation factors are defined as

\begin{equation}
\alpha^C = \sum_{g\in\mathcal{G}} c_g,
\end{equation}
\begin{equation}
\alpha^P = \gcd\left(\{P_g\}_{g\in\mathcal{G}},\,D\right)
\end{equation}
where the factor $\alpha^C$ normalises the cost term, while $\alpha^P$ expresses the demand-balance mismatch in the smallest common discrete power unit of the instance, measured by the greatest common divisor between the powers offered by the generation units and the total demand. The corresponding normalised coefficients are
\begin{equation}
\tilde{c}_g = \frac{c_g}{\alpha^C},
\qquad
\tilde{P}_g = \frac{P_g}{\alpha^P},
\qquad
\tilde{D}_t = \frac{D_t}{\alpha^P}.
\end{equation}

The effective penalty coefficient is constructed in two stages. First, the minimum penalty value required to preserve the optimal feasible solution as the ground state of the QUBO is denoted by $\lambda_{\min}$. An automatic instance-dependent baseline is then defined as

\begin{equation}
\lambda_{\mathrm{auto}}
=
2\lambda_{\min}
\left(
1+\frac{1}{N}
\right),
\end{equation}

where $N=|\mathcal{G}|$ is the number of binary variables, and therefore the number of logical qubits. The factor $(1+1/N)$ introduces a small dimension-dependent safety margin. Its effect is slightly larger for low-dimensional instances, where the penalty calibration is more sensitive, and progressively decreases as the number of qubits increases, avoiding excessive penalisation in larger problems.

Then, a dimensionless multiplier $\eta$ is introduced, yielding

\begin{equation}
\lambda = \eta\,\lambda_{\mathrm{auto}}.
\end{equation}

The same value of $\eta$ is used for all UC instances. In this work, $\eta=0.5$, and therefore the effective penalty becomes

\begin{equation}
\lambda
=
\lambda_{\min}
\left(
1+\frac{1}{N}
\right).
\end{equation}

This construction provides a penalty coefficient with a consistent interpretation across UC instances with different power and cost scales, while introducing a smooth correction with the dimensionality of the problem.

Using these definitions, the normalised QUBO form for a commitment-based UC instance can be written as

\begin{equation}
\min_{\mathbf{u_{g}}}
\left[
\sum_{g\in\mathcal{G}} \tilde{c}_g u_{g}
+
\lambda
\left(
\sum_{g\in\mathcal{G}} \tilde{P}_g u_{g}
-
\tilde{D}
\right)^2
\right].
\end{equation}
    
%

Once the QUBO formulation is obtained, one step is needed before changing the problem to a \textit{quantum form}. A change of variables is applied to express the problem in \textit{spin form},
\begin{equation}
z_g = 1-2u_g,
\qquad
u_g\in\{0,1\},
\qquad
z_g\in\{+1,-1\}.
\end{equation}

This transformation leads to the following minimisation problem:
\begin{equation}\label{eq:normalised_qubo}
\min_{\mathbf{z}_{g}}
\left[
\sum_{g\in\mathcal{G}} \tilde{c}_g \left(\frac{1-z_g}{2}\right)
+
\lambda
\left(
\sum_{g\in\mathcal{G}} \tilde{P}_g \left(\frac{1-z_g}{2}\right)
-
\tilde{D}
\right)^2
\right].
\end{equation}
To obtain the corresponding quantum formulation, the classical spin variables
\(z_g\) are promoted to Pauli-\(Z\) operators \(Z_g\),
\begin{equation}
    z_{g} \longrightarrow Z_{g},
    \qquad
    Z_{g} =
    \begin{pmatrix}
    1 & 0 \\
    0 & -1
    \end{pmatrix},
\end{equation}
whose eigenvalues in the computational basis are \(\{+1,-1\}\). Since one binary commitment variable is assigned to each generating unit, the number of
logical qubits is equal to the number of generators, i.e.,
\begin{equation}
    N = |\mathcal{G}|.
\end{equation}
Therefore, an arbitrary but fixed ordering of the generating units can be
introduced as
\(\mathcal{G}=\{g_1,g_2,\ldots,g_N\}\). Under this ordering, each generator
spin operator \(Z_{g}\) can be relabelled as a logical qubit operator,
\begin{equation}
    Z_i \equiv Z_{g_i},
    \qquad i=1,\ldots,N.
\end{equation}
After expanding the quadratic penalty term in
\eqref{eq:normalised_qubo}, collecting linear, quadratic, and constant
contributions, and replacing the classical spin variables by the corresponding
operators, the resulting Hamiltonian is formally equivalent to an Ising
Hamiltonian:
\begin{equation}\label{eq_ising_hamiltonian}
H_{\text{Ising}}
=
\sum_{i=1}^{N} h_i Z_i
+
\sum_{i<j} J_{i,j} Z_i Z_j
+
E_0 ,
\end{equation}
where \(h_i\) contains the coefficients of the linear spin terms, \(J_{i,j}\)
contains the pairwise interaction coefficients generated by the quadratic
penalty, and \(E_0\) is a constant energy offset. The indices \(i\) and \(j\)
therefore denote logical qubits associated with the ordered generating units. This Hamiltonian is in the required form to be used as the problem Hamiltonian in the subsequent adiabatic quantum evolution, unless the optional PCE compression stage is applied.

\subsection{PCE-based qubit reduction}
\label{subsec:PCE_methodology}

Pauli Correlation Encoding (PCE) is used in this work as a qubit-compression pre-processing stage. It constructs a reduced Hamiltonian acting on $m<N$ effective qubits from an original Ising Hamiltonian defined on $N$ logical spin variables. The objective is not to obtain a lossless representation of the complete $2^N$ search space, but to restrict the quantum evolution to a structured subset of logical configurations generated by correlations among the $m$ reduced spins. Therefore, PCE introduces a compressed representation trade-off where the problem dimension is reduced from $2^N$ to at most $2^m$, but only the logical solutions representable by the selected encoding can be recovered.

In this work, we use a simple $Z$ reduction PCE configuration, consistent with the implemented Hamiltonian representation. Each original logical spin operator \(Z_i\), associated with the logical spin value \(z_i\), is replaced by a product of Pauli-\(Z\) operators acting on the reduced qubit register. This produces a reduced Hamiltonian that remains diagonal in the computational basis and can therefore be used within the same adiabatic-evolution framework as the original Ising formulation.

Let the original Ising Hamiltonian ($H_{\mathrm{Ising}}$) be defined over $N$ logical spin variables. PCE introduces a reduced register of $m<N$ spin variables,
\begin{equation}
\mathbf{s} = (s_0,\ldots,s_{m-1}),
\qquad
s_q\in\{-1,+1\}.
\end{equation}
The value of $m$ controls the compression level. Smaller values of $m$ provide stronger qubit reduction but also restrict the number of original UC schedules that can be represented after decoding.

Instead of assigning each original logical spin to one reduced spin, PCE assigns each logical spin $Z_i$ to a Pauli-$Z$ string over a support set $S_i\subseteq\{0,\ldots,m-1\}$:
\begin{equation}
Z_i
\;\longmapsto\;
\Pi_i
=
\prod_{q\in S_i} Z_q.
\end{equation}
Equivalently, for a reduced computational-basis assignment $\mathbf{s}\in\{-1,+1\}^{m}$, the decoded logical spin is
\begin{equation}
\hat{z}_i
=
\prod_{q\in S_i} s_q,
\qquad
i=1,\ldots,N.
\end{equation}
Thus, PCE defines a mapping
\begin{equation}
f_{\mathrm{PCE}}:
\{-1,+1\}^{m}
\rightarrow
\{-1,+1\}^{N},
\end{equation}
from reduced spin assignments to logical assignments in the original UC space.

The support size $|S_i|$ is controlled by the selected correlation orders. In the implementation, these orders are defined through the parameter \texttt{k}. For example, \texttt{k=(1,2)} means that each logical spin can be encoded either as a single reduced $Z$ operator or as a product of two reduced $Z$ operators. This parameter controls the expressiveness of the encoding, although it does not directly impose the locality of the final reduced Hamiltonian.

Given the original Ising Hamiltonian (Eq. \ref{eq_ising_hamiltonian}),
the PCE-reduced Hamiltonian is obtained by substituting each logical operator $Z_i$ with its encoded Pauli string $\Pi_i$:
\begin{equation}
H_{\mathrm{PCE}}
=
H(\Pi_1,\ldots,\Pi_N).
\end{equation}
After this substitution, a linear term $h_i Z_i$ becomes the Pauli string $h_i\Pi_i$, while a quadratic term $J_{ij}Z_iZ_j$ becomes
\begin{equation}
J_{ij}Z_iZ_j
\;\longmapsto\;
J_{i,j}\Pi_i\Pi_j.
\end{equation}

Therefore, even if the original Ising Hamiltonian is quadratic, the PCE substitution can generate higher-order Pauli-\(Z\) products in the reduced Hamiltonian. This is, nevertheless, a direct consequence of encoding each logical spin as a correlation of reduced qubits. In this sense, PCE can transform the original QUBO Ising model into an effective HUBO (High-order Unconstrained Binary Optimisation) reduced formulation on fewer variables. The optimisation is performed in this compressed encoded space, and the resulting reduced bitstrings are then decoded back into assignments of the original UC variables. Thus, the interaction order may increase during the reduction, but the recovered candidate solutions are still interpreted in the original problem space.

In the implemented UC workflow, the support sets \(S_i\) are not imposed manually. Instead, they are selected through a demand-shell-guided search. Since the system demand is known beforehand, the demand-balance constraint provides a natural way of identifying feasible configurations that are relevant to the compressed representation. The feasible configurations contained in the target shell are then weighted according to their generation cost during the encoding-selection stage, so that candidate encodings are evaluated not only by the feasible states they preserve, but also by the relevance of those states for the UC objective. This weighting is used only to select the compressed representation; the subsequent adiabatic evolution is performed on the resulting PCE Hamiltonian.

The feasible configurations in the target shell are assigned an objective-informed weight according to their generation cost,
\begin{equation}
w(\mathbf{u})
\propto
\exp\left[-\lambda_{\mathrm{cost}} C(\mathbf{u})\right],
\end{equation}
where \(C(\mathbf{u})\) is the original UC cost and \(\lambda_{\mathrm{cost}}\) controls the strength of the prioritisation. Candidate PCE encodings are evaluated according to the weighted fraction of the target shell that remains representable. In this way, the search gives preference to compressed representations that preserve low-cost feasible regions, reducing the chance that relevant schedules are discarded during qubit reduction. The weighting is used only to select the PCE encoding; once the encoding is fixed, the adiabatic evolution is performed on the resulting reduced Hamiltonian obtained by substituting \(Z_i\mapsto \Pi_i\).


\subsection{Adiabatic quantum evolution and sampling}
\label{subsec:adiabatic_solution}

Once the problem Hamiltonian has been constructed, the next stage consists of solving the optimisation problem through adiabatic quantum evolution.
\begin{equation}
    |\psi\rangle^0 \underbrace{\longrightarrow}_{AE} |\psi\rangle^{opt}
\end{equation}
If PCE is not applied, the evolution is performed on the original $N$-qubit Ising Hamiltonian . If PCE is applied, the evolution is performed on the reduced $m$-qubit Hamiltonian $H^{\mathrm{PCE}}$. For generality, let $n_{\mathrm{eff}}$ denote the number of qubits of the space where the quantum evolution is executed:
\begin{equation}
n_{\mathrm{eff}}
= 
\begin{cases}
N, & \text{without PCE},\\
m, & \text{with PCE}.
\end{cases}
\end{equation}
Accordingly, the problem Hamiltonian used by the quantum solver is
\begin{equation}
H_{\mathrm{problem}}
=
\begin{cases}
H_{Ising}, & \text{without PCE},\\
H_{\mathrm{PCE}}, & \text{with PCE}.
\end{cases}
\end{equation}

The optimal solution is defined by the eigenstate related with the lowest eigenvalue of $H_{problem}$. To get this state at the end of the process, it is necessary to define an \textit{easy} initial Hamiltonian whose lowest-value state is known. This \textit{easy} Hamiltonian is defined as:
\begin{equation} 
H_{\mathrm{easy}} = -\sum_{q=1}^{n_{\mathrm{eff}}} X_q 
\quad, \quad
X = \begin{pmatrix}
0 & 1 \\
1 & 0
\end{pmatrix}. 
\end{equation} 
whose lowest-energy state is defined as:
\begin{equation}
|\psi\rangle^0
=
\frac{1}{\sqrt{2^{n_{\mathrm{eff}}}}}
\sum_{b\in\{0,1\}^{n_{\mathrm{eff}}}}
|b\rangle 
\end{equation}
where $|b\rangle$ is a bitstring of the computational basis.

$H_0$ is also known as \textit{driver} Hamiltonian because it is interpolated with the problem Hamiltonian to transform this \textit{easy} initial state to the final optimal one through the adiabatic path:
\begin{equation}
H(t)
=
\left(1-\frac{t}{T} \right)H_{\mathrm{easy}}
+
\frac{t}{T} H_{\mathrm{problem}}
\end{equation}
where the boundary conditions are
\begin{equation}
    H(0) = H_{\mathrm{easy}} \quad, \quad H(T)=H_\mathrm{problem}.
\end{equation}
The continuous evolution is numerically approximated using discrete evolution steps controlled by the final evolution time $T$ and the time step $\Delta t$. These parameters affect both the quality of the adiabatic approximation and the computational cost of the simulation. In this work, the evolution is implemented using Qibo, and the same evolution settings are used when comparing original and PCE-compressed Hamiltonians in order to isolate the effect of the encoding.

After the evolution, the final quantum state is measured repeatedly in the computational basis. This produces a frequency distribution over measured bitstrings,
\begin{equation}
f(b),
\qquad
b\in\{0,1\}^{n_{\mathrm{eff}}}.
\end{equation}
The most probable bitstrings represent the candidate solutions generated by the quantum evolution. The measured distribution can be passed directly to the decoding stage, or it can be refined first using the optional SQD post-processing stage described in Subsection~\ref{subsec:sqd_methodology}.

\subsection{SQD post-processing}
\label{subsec:sqd_methodology}

Sample-based Quantum Diagonalisation (SQD) is incorporated as an optional hybrid post-processing stage. Its role is to refine the information already obtained from the quantum evolution without requiring a new quantum simulation. Rather than exploring the full Hilbert space again, SQD builds a reduced subspace using the bitstrings observed in the measurements, projects an auxiliary Hamiltonian onto that subspace, and diagonalises it classically.

If PCE is used, SQD acts on the reduced $m$-qubit space. Otherwise, it acts on the original $N$-qubit space. Therefore, SQD is always defined over the same effective space where the adiabatic quantum evolution has been executed.

Let the set of distinct observed states be
\begin{equation}
\Omega
=
\left\{
b\in\{0,1\}^{n_{\mathrm{eff}}}
:
f(b)>0
\right\}.
\end{equation}
The full effective Hilbert-space dimension is
\begin{equation}
d = 2^{n_{\mathrm{eff}}},
\end{equation}
and the number of different measured bitstrings is
\begin{equation}
U = |\Omega|.
\end{equation}
The sampled subspace is built from the most relevant measured bitstrings using a
single target retained fraction, \(\rho_{\mathrm{SQD}}\). This parameter directly
sets the desired SQD subspace size relative to the full effective Hilbert space:
\begin{equation}
K_{\mathrm{target}}
=
\rho_{\mathrm{SQD}} d
 .
\end{equation}
The measured bitstrings are then sorted in decreasing order of frequency, and the SQD subspace is defined by retaining the most frequent measured configurations:
\begin{equation}
S
=
\left\{
b^{(1)},b^{(2)},\ldots,b^{(M)}
\right\},
\quad
M
=
\min\left(
K_{\mathrm{target}},
U
\right).
\end{equation}
Therefore, the SQD subspace is always constructed only from states that were
actually observed in the measurement stage. The target retained fraction
\(\rho_{\mathrm{SQD}}\) controls the intended dimensional reduction, while the realised retained fraction is
\begin{equation}
\rho_{\mathrm{eff}}
=
\frac{|S|}{d}
=
\frac{M}{d}.
\end{equation}
Equivalently, the realised SQD reduction is
\begin{equation}
r_{\mathrm{SQD}}
=
1-\rho_{\mathrm{eff}}.
\end{equation}
If the adiabatic evolution samples fewer distinct states than the target SQD
dimension, the subspace is naturally capped by the observed set \(\Omega\). This
is not an additional user-controlled reduction parameter, but a consequence of
the sample-guided construction of the SQD subspace. This strategy restricts the
refinement to the most representative sampled configurations while keeping the
classical diagonalisation tractable.

Once the sampled subspace has been selected, SQD constructs an auxiliary Hamiltonian
\begin{equation}
H^{*}(\beta)
=
(1-\beta)
\sum_{i=1}^{n_{\mathrm{eff}}}X_i
+
\beta H_{\mathrm{problem}},
\end{equation}
where $H^{\mathrm{problem}}$ is the Ising Hamiltonian used in the adiabatic evolution and $\beta\in(0,1]$ controls the balance between the transverse-field contribution and the original problem Hamiltonian. Values of $\beta$ close to one favour the low-energy sampled states according to the original problem Hamiltonian, whereas smaller values increase the influence of one-bit-flip couplings between sampled configurations.

The auxiliary Hamiltonian is projected onto the sampled subspace, producing a reduced matrix
\begin{equation}
H_{\mathrm{sub}}
\in
\mathbb{R}^{M\times M}.
\end{equation}
The diagonal entries are obtained from the problem energy of each sampled state:
\begin{equation}
\left[H_{\mathrm{sub}}\right]_{j,j}
=
\beta
\left\langle
b^{(j)}
\middle|
H_{\mathrm{problem}}
\middle|
b^{(j)}
\right\rangle.
\end{equation}
The off-diagonal entries come from the transverse-field contribution. Since a Pauli-$X$ operator flips one qubit, two sampled bitstrings are coupled only if
they differ in exactly one bit. In that case, the coupling weight is $(1-\beta)$. As a result, the projected matrix preserves both the energy of each sampled state and the local one-bit-flip connectivity between sampled configurations.

Finally, SQD solves the reduced eigenvalue problem
\begin{equation}
H_{\mathrm{sub}}v_0=e_0v_0,
\end{equation}
and keeps its ground-state eigenpair. The obtained eigenvector provides a refined amplitude distribution over the sampled states contained in $S$, defining the SQD-refined state
\begin{equation}
|\Psi\rangle^{\mathrm{SQD}}
=
\sum_{j=1}^{M}
v_{0,j}
\,
|b^{(j)}\rangle.
\end{equation}
SQD does not modify the optimisation problem itself and cannot introduce configurations outside the sampled subspace $S$. Its quality therefore depends on the ability of the adiabatic evolution and measurement stage to generate representative candidate states. When the sampled set already contains feasible or near-optimal configurations, SQD provides a low-cost mechanism to sharpen the final probability distribution before decoding.

\subsection{Decode and recover UC solutions}
\label{subsec:decoding_uc}

The decoding stage is needed in all configurations, with or without SQD. Its role is to translate the output of the quantum procedure into binary UC variables that can be evaluated in the original problem space. Depending on the workflow, the input to this stage is either the measured distribution obtained after the adiabatic evolution or the probability distribution refined by SQD.

When SQD is not used, the probability assigned to each measured bitstring is obtained directly from its sampling frequency:
\begin{equation}
P_{\mathrm{eff}}(b)
=
\frac{f(b)}
{\sum_{b'} f(b')}.
\end{equation}
When SQD is applied, the probability of each retained sampled state is instead computed from the refined eigenvector amplitudes:
\begin{equation}
P_{\mathrm{eff}}\left(b^{(j)}\right)
=
\left|v_{0,j}\right|^2,
\qquad
b^{(j)}\in S.
\end{equation}

If PCE is not applied, the measured bitstrings already represent the original logical variables of the UC formulation. In this case, the corresponding spin variables are recovered directly from the binary variables as
\begin{equation}
z_i = 1-2x_i.
\end{equation}

When PCE is used, the measured or SQD-refined states belong to the reduced \(m\)-qubit space. Therefore, each reduced bitstring is first converted into a reduced spin assignment \(\mathbf{s}\), and the original logical spins are then reconstructed through the PCE correlation map:
\begin{equation}
\hat{z}_i = \prod_{q\in S_i} s_q, \qquad i=1,\ldots,N.
\end{equation}
The decoded spins are finally converted back into binary UC variables:
\begin{equation}
\hat{x}_i = \frac{1-\hat{z}_i}{2}.
\end{equation}

The resulting binary vector \(\hat{\mathbf{x}}\) is interpreted as a UC commitment schedule. In the commitment-based formulation considered here, this corresponds to the on/off decisions \(\hat{u}_{g,t}\) of the generators for the analysed time step or time horizon. Since different reduced states may decode into the same original UC schedule, their probabilities are aggregated:

\begin{equation}
P_{\mathrm{UC}}(\hat{\mathbf{x}}) = \sum_{b:\,\mathcal{D}(b)=\hat{\mathbf{x}}} P_{\mathrm{eff}}(b),
\end{equation}

where \(\mathcal{D}(\cdot)\) denotes the decoding map from the effective quantum space to the original binary UC space. This aggregation is particularly relevant when PCE is active, because the reduced representation may assign several effective states to the same decoded schedule.

The output of this stage is therefore a probability distribution over UC schedules expressed in the original decision-variable space. These decoded schedules are then used in the final evaluation step.

After decoding the results, the last stage evaluates the decoded schedules using the original UC formulation. This evaluation is independent of how the candidate solutions were obtained: direct adiabatic evolution, PCE preprocessing, SQD post-processing, or the combined PCE+SQD workflow. In all cases, feasibility and optimality are assessed with respect to the original UC objective and constraints, rather than only through the Hamiltonian explored during the quantum evolution.

For a decoded schedule \(\hat{\mathbf{u}}\), the demand-balance residual at time \(t\) is computed as
\begin{equation}
r(\hat{\mathbf{u}}) = \sum_{g\in\mathcal{G}} P_g \hat{u}_{g} - D.
\end{equation}
A schedule is considered feasible when this residual satisfies the demand-balance condition and the remaining operational constraints are fulfilled. In the simplified commitment-based instances used in this work, feasibility is mainly determined by satisfying the power-balance condition at the natural discrete resolution of the problem.

Each decoded schedule is also evaluated according to the original UC cost function. For the commitment-based formulation, this cost is given by
\begin{equation}
C(\hat{\mathbf{u}}) = \sum_{g\in\mathcal{G}} c_g \hat{u}_{g},
\end{equation}
or by the corresponding generation-cost expression when dispatch variables or discretised production levels are explicitly included.

The total probability assigned to feasible schedules is then computed as
\begin{equation}
P_{\mathrm{feas}} = \sum_{\hat{\mathbf{x}}\in\mathcal{F}} P_{\mathrm{UC}} (\hat{\mathbf{x}}),
\end{equation}
where \(\mathcal{F}\) denotes the set of feasible decoded UC schedules. Similarly, if \(\mathcal{O}\) is the set of optimal or best-known schedules, the probability assigned to optimal solutions is
\begin{equation}
P_{\mathrm{opt}} = \sum_{\hat{\mathbf{x}}\in\mathcal{O}} P_{\mathrm{UC}} (\hat{\mathbf{x}}).
\end{equation}

The resource advantage provided by PCE is quantified through the relative qubit reduction. If the original Ising formulation requires \(N\) logical qubits and the quantum evolution is performed with \(n_{\mathrm{eff}}\) effective qubits, the reduction is

\begin{equation}
R_q
=
1
-
\frac{n_{\mathrm{eff}}}{N}.
\end{equation}
For unreduced runs, \(n_{\mathrm{eff}}=N\), and therefore \(R_q=0\). For PCE-compressed runs, \(n_{\mathrm{eff}}=m<N\), which leads to a positive reduction.

In addition to qubit reduction, the computational cost of each workflow is quantified through the total measured runtime. This KPI includes the quantum-evolution time and, when applicable, the extra time required by the PCE construction and SQD post-processing stages:

\begin{equation}
T_{\mathrm{tot}}
=
\,T_{\mathrm{PCE}}
+
T_{\mathrm{AE}}
+
\,T_{\mathrm{SQD}},
\end{equation}
where \(T_{\mathrm{AE}}\) is the adiabatic-evolution time, \(T_{\mathrm{PCE}}\) is the PCE preprocessing time, and \(T_{\mathrm{SQD}}\) is the SQD post-processing time. This definition allows the direct AE, PCE-only, SQD-only, and PCE+SQD configurations to be compared using the same time metric.

The final evaluation combines three complementary aspects: feasibility and objective value in the original UC problem, probability assigned to feasible and optimal schedules, and the qubit reduction achieved by the encoding. Using the same evaluation criteria for all configurations makes it possible to compare the baseline, PCE-only, SQD-only and PCE + SQD workflows under a common framework.

\begin{algorithm}[!t]
\caption{Hybrid quantum UC workflow with optional PCE and SQD}
\label{alg:quantum_uc_workflow}
\footnotesize
\begin{algorithmic}[1]

\STATE \textbf{Input:} UC data $\{c_g,P_g,D_t\}$; penalty parameter $\eta$; AQE parameters $T$, $\Delta t$, $\tau$; optional PCE parameters $(m,\texttt{k\_list})$; optional SQD parameters $(\rho_{\mathrm{SQD}},\beta)$.
\STATE \textbf{Output:} Decoded distribution $P_{\mathrm{UC}}(\hat{\mathbf{x}})$, best feasible schedule $\hat{\mathbf{u}}^{\star}$, and metrics $P_{\mathrm{feas}}$, $P_{\mathrm{opt}}$, $R_q$.

\medskip
\STATE \textbf{UC-to-Ising mapping:} normalise the UC data, set $\lambda=\eta\lambda_{\mathrm{auto}}$, build the QUBO, and map binary variables to spins through $z_i=1-2x_i$.
\STATE Obtain the Ising Hamiltonian
$H_1=\sum_i h_iZ_i+\sum_{i<j}J_{ij}Z_iZ_j+E_0$.

\medskip
\IF{\textsc{pce}}
    \STATE Build the demand-guided feasible shell, apply the cost-based weighting, select supports $S_i$ from $m$ and \texttt{k\_list}, and apply $Z_i\mapsto \Pi_i=\prod_{q\in S_i}Z_q$.
    \STATE Set $H_{\mathrm{problem}}\leftarrow H_1^{\mathrm{PCE}}$ and $n_{\mathrm{eff}}\leftarrow m$.
\ELSE
    \STATE Set $H_{\mathrm{problem}}\leftarrow H_1$ and $n_{\mathrm{eff}}\leftarrow N$.
\ENDIF

\medskip
\STATE \textbf{AQE and sampling:} evolve the initial equal-superposition state towards $H_{\mathrm{problem}}$ through $H(t)$ using $T$, $\Delta t$, and $\tau$.
\STATE Measure the final state and obtain bitstring frequencies $f(b)$, with $b\in\{0,1\}^{n_{\mathrm{eff}}}$.

\medskip
\IF{\textsc{sqd}}
    \STATE Retain the most frequent sampled states in $\mathcal{S}_{\mathrm{SQD}}$ according to $\rho_{\mathrm{SQD}}$.
    \STATE Project $H^{*}(\beta)$ onto $\mathcal{S}_{\mathrm{SQD}}$, diagonalise $H_{\mathrm{sub}}$, and set $P_{\mathrm{eff}}(b^{(j)})=|v_{0,j}|^2$.
\ELSE
    \STATE Set $P_{\mathrm{eff}}(b)=f(b)/\sum_{b'}f(b')$.
\ENDIF

\medskip
\STATE \textbf{Decoding:} for each sampled state, decode it into the original UC space. If PCE is active, use $\hat z_i=\prod_{q\in S_i}s_q$; otherwise decode directly.
\STATE Convert decoded spins into binary schedules $\hat{\mathbf{x}}$, aggregate repeated decoded states into $P_{\mathrm{UC}}(\hat{\mathbf{x}})$.

\medskip
\STATE \textbf{Evaluation:} evaluate all decoded schedules with the original UC cost and feasibility constraints.
\STATE Compute $P_{\mathrm{feas}}$, $P_{\mathrm{opt}}$, and $R_q=1-n_{\mathrm{eff}}/N$.
\STATE Select $\hat{\mathbf{u}}^{\star}$ as the best feasible decoded schedule.

\end{algorithmic}
\end{algorithm}



\section{Case Studies and results}

This section evaluates the proposed pre- and post-processing techniques for the UC problem across different system sizes. PCE and SQD are first analysed independently in Sections~\ref{subsec:PCE_Results} and~\ref{subsec:SQD_Results}, respectively, before both techniques are combined in Section~\ref{subsec:PCE_SQD_Results}. The objective is to evaluate their impact on KPI (Key Performance Indicators) such as scalability, computational cost, feasible-state probability, and optimal-solution probability as the number of generators, and consequently the number of qubits, increases.

The analysed UC instances range from \(N=4\) to \(N=20\) generators, with each case built on the same base system by progressively adding new generating units. Each generator is represented by one binary commitment variable and is characterised by a fixed available power \(P_g\) and a corresponding commitment cost. The power values are integer quantities expressed in MW and remain within the range considered for the constructed case-study family. For each instance, the commitment vector must satisfy the power balance constraint
\begin{equation}
\begin{aligned}
\min_{\mathbf{u}_g} \quad
& \sum_{g\in\mathcal{G}} c_g u_g \\
s.t. \quad
& \sum_{g\in\mathcal{G}} P_g u_g = D.
\end{aligned}
\end{equation}
where \(u_g \in \{0,1\}\) represents the commitment status of generator \(g\) and \(D\) is the system demand.

The case study family is constructed progressively, so that larger instances extend the smaller ones by adding new generators while preserving the previously defined units. This provides a controlled benchmark in which the number of binary variables, and consequently the original Hilbert-space dimension, increases with the system size, allowing the proposed techniques to be evaluated under progressively larger UC instances.

Unless otherwise stated, all adiabatic simulations are performed with time step \(\Delta t=0.1\), and  total evolution time \(T=100\) for comparing the scalability on the cases with reliable results and no much time spent in the evolution. For the  normalised UC Hamiltonian, the penalty term is parameterised as $\lambda =  \eta\lambda_{\mathrm{auto}}$, where \(\lambda_{\mathrm{auto}}\) denotes the instance-dependent penalty scaling introduced during the UC to QUBO transformation and \(\eta\) is a dimensionless multiplier, set as \(\eta=0.5\), used for the experiments reported in this
section in order to maintain the same penalty configuration across the analysed case-study family. For PCE, the correlation orders are fixed to \(\texttt{k}=(1,2,3,4)\), using 3000 trials per reduced dimension with a fixed random seed of 0. The demand shell contains the feasible UC configurations and uses a cost-prioritisation factor of \(\lambda_{\mathrm{cost}}=0.20\). The reduced dimension \(m\) is varied during the calibration study and subsequently fixed to the selected value for each system size.

The direct AE and AE+SQD configurations are evaluated up to \(N=12\), since the classical state-vector simulation of the original Hamiltonian becomes increasingly demanding as \(2^N\) grows. PCE-based approaches are instead evaluated up to \(N=20\), since the original \(N\)-qubit problem is compressed into an \(m\)-qubit Hamiltonian before the adiabatic evolution.


\subsection{PCE based UC}
\label{subsec:PCE_Results}

As described in Section~\ref{subsec:PCE_methodology}, PCE compresses the original \(N\)-qubit Ising Hamiltonian into a reduced representation acting on \(m<N\) effective qubits. The purpose of this reduction is to decrease the Hilbert-space dimension evolved during the adiabatic simulation while preserving a subset of configurations that contains relevant feasible and high quality UC schedules. The performance of PCE depends on the selected reduced dimension \(m\) and on the encoding constructed for that dimension. For this reason, several values of \(m\) are evaluated for each original system size. 

Table~\ref{tab:PCE_summary} summarises the selected PCE configuration for each UC instance overall the possible values from: 
\begin{equation}
    \frac{N}{2}\leq m \leq N-1 \text{ or } 12
\end{equation}
The best reduced dimension is selected according to the highest probability assigned to the optimal or best feasible decoded schedule, denoted by \(P_{\mathrm{opt}}\). The table also reports the total feasible probability \(P_{\mathrm{feas}}\), the feasible state coverage \(C_{\mathrm{feas}}\), the PCE pre-processing time, and the evolution time of the selected reduced Hamiltonian.

\begin{table*}[h!]
\centering
\scriptsize
\renewcommand{\arraystretch}{1.05}
\resizebox{\textwidth}{!}{%
\begin{tabular}{c c c c c c c c c c}
\hline
\(N\) &
Feasible &
Tested \(m\) &
Best \(m\) &
Red. (\%) &
PCE T (s) &
Evol. T (s) &
\(P_{\mathrm{opt}}\) &
\(P_{\mathrm{feas}}\) &
\(C_{\mathrm{feas}}\) \\
\hline
4  & 3  & $\{2,3\}$                  & 2  & 50.0 & 0.89   & 5.81    & 0.778 & 0.977 & $2/3$  \\
5  & 4  & $\{2,3,4\}$                & 2  & 60.0 & 6.10   & 12.13   & 0.807 & 1.000 & $2/4$  \\
6  & 5  & $\{3,4,5\}$                & 3  & 50.0 & 2.08   & 14.07   & 0.414 & 0.935 & $3/5$  \\
8  & 9  & $\{4,5,6,7\}$              & 4  & 50.0 & 5.36   & 28.56   & 0.343 & 0.539 & $4/9$  \\
10 & 11 & $\{5,6,7,8,9\}$            & 6  & 40.0 & 8.28   & 107.40  & 0.591 & 0.905 & $4/11$ \\
12 & 13 & $\{6,7,8,9,10,11,12\}$     & 6  & 50.0 & 42.01  & 297.50  & 0.625 & 0.771 & $2/13$ \\
14 & 16 & $\{7,8,9,10,11,12\}$       & 8  & 42.9 & 60.57  & 474.71  & 0.709 & 0.944 & $3/16$ \\
16 & 20 & $\{8,9,10,11,12\}$         & 8  & 50.0 & 25.03  & 387.92  & 0.590 & 0.915 & $2/20$ \\
18 & 24 & $\{9,10,11,12\}$           & 9  & 50.0 & 35.23  & 559.74  & 0.890 & 0.999 & $2/24$ \\
20 & 28 & $\{10,11,12\}$             & 10 & 50.0 & 123.86 & 1067.77 & 0.906 & 0.906 & $2/28$ \\
\hline
\end{tabular}
}
\vspace{1mm}
\caption{Summary of the selected PCE configurations for the UC case studies. The column ``Feasible'' reports the total number of feasible configurations of each original UC instance. The PCE time corresponds to the pre-processing time required to construct the reduced model, while the evolution time includes circuit construction and execution. \(C_{\mathrm{feas}}\) denotes the number of feasible configurations represented by the selected PCE encoding relative to the total number of feasible configurations of the original UC problem.}
\label{tab:PCE_summary}
\end{table*}


The results show that PCE can significantly reduce the number of evolved qubits while still preserving useful feasible and optimal configurations. The selected reduction ranges from \(40\%\) to \(60\%\), depending on the instance. However, the results also indicate that the relationship between compression and solution quality is not linear. A larger reduction does not necessarily imply a poorer probability distribution, and conversely, a relatively moderate reduction does not guarantee high optimal probability.

\begin{figure}[!ht]
\centering
\includegraphics[width=1\linewidth]{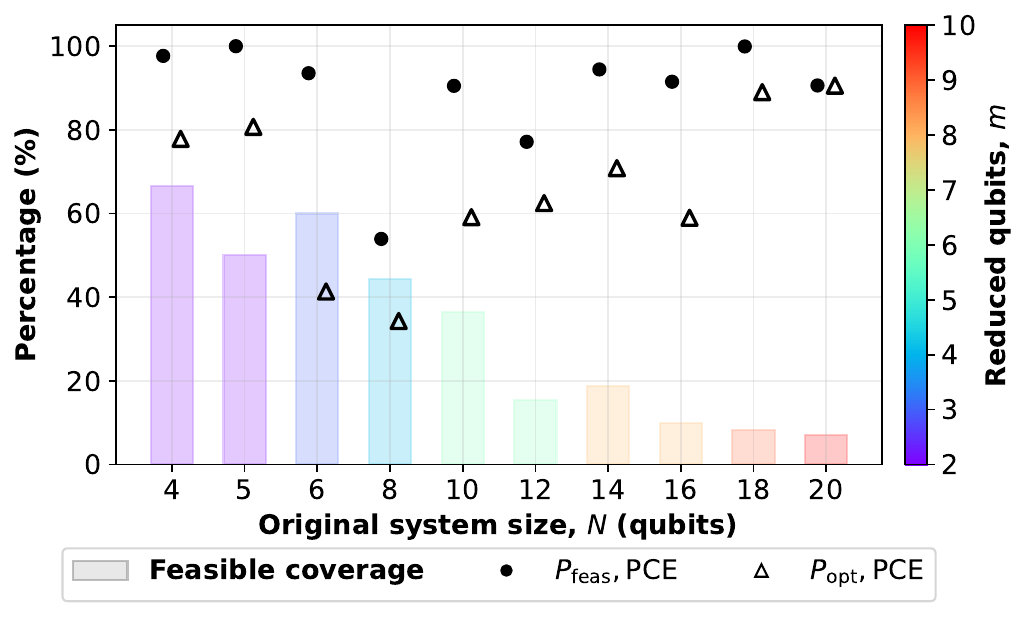}
\caption{Best PCE configuration for each UC case study. Bars represent the feasible-state coverage \(C_{\mathrm{feas}}\) of the selected reduced representation. Circular markers show the total feasible probability \(P_{\mathrm{feas}}\), while triangular markers show the optimal or best feasible probability \(P_{\mathrm{opt}}\). The colour scale indicates the reduced number of qubits \(m\).}
\label{fig:PCE_best_summary}
\end{figure}

Figure~\ref{fig:PCE_best_summary} provides a visual summary of the selected configurations. The bars represent the fraction of feasible schedules preserved by the selected reduced representation, while the markers show the feasible and optimal probabilities obtained after PCE-based adiabatic evolution. The colour scale represents the selected reduced dimension \(m\).

Overall, PCE achieves substantial dimensionality reductions, while maintaining high feasible and optimal probabilities for most UC instances. The results also show that feasible-state coverage alone does not determine performance: for example, the \(N=18\) and \(N=20\) cases retain only \(2/24\) and \(2/28\) feasible configurations, respectively, yet achieve \(P_{\mathrm{opt}}=0.890\) and \(0.906\). Conversely, lower performance is observed for \(N=8\), despite retaining \(4/9\) feasible states. This indicates that PCE performance depends more on preserving the most relevant feasible configurations and on the resulting reduced Hamiltonian than on maximising feasible state coverage itself.


\subsection{SQD UC}
\label{subsec:SQD_Results}

This subsection evaluates SQD as a post-processing stage applied to the direct UC adiabatic simulations. Unlike PCE, SQD does not modify the Hamiltonian before the quantum evolution, but instead refines the measured distribution by constructing and diagonalising a reduced sampled subspace.

The selected SQD operating point uses \(\beta=1.0\) and a target retained fraction $\rho_{\mathrm{SQD}}=\frac{1}{8}.$ The realised subspace size can differ from this nominal value depending on the number of distinct sampled configurations. Table~\ref{tab:SQD_summary} summarises the SQD results for the direct UC cases from \(N=4\) to \(N=12\).

\begin{figure}[!ht]
    \centering
    \includegraphics[width=1\linewidth]{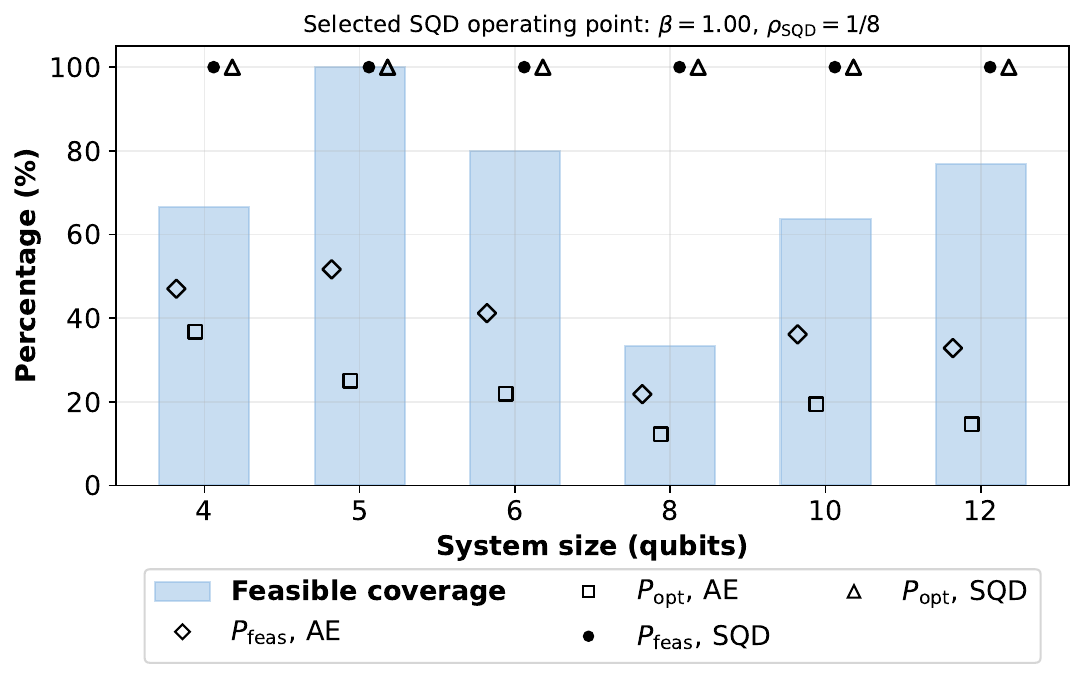}
    \caption{Selected SQD operating point for the direct UC case studies with \(\beta=1.0\) and \(\rho_{\mathrm{SQD}}=1/8\). Bars show the feasible-state coverage \(C_{\mathrm{feas}}\). Markers show the feasible and optimal probabilities before SQD, \(P_{\mathrm{feas}}^{\mathrm{AE}}\) and \(P_{\mathrm{opt}}^{\mathrm{AE}}\), and after SQD, \(P_{\mathrm{feas}}^{\mathrm{SQD}}\) and \(P_{\mathrm{opt}}^{\mathrm{SQD}}\).}
    \label{fig:sqd_best_beta_rho}
\end{figure}

Figure~\ref{fig:sqd_best_beta_rho} visualises the same operating point, comparing the feasible and optimal probabilities before and after SQD.

SQD consistently concentrates the sampled distribution on the optimal schedule: \(P_{\mathrm{opt}}^{\mathrm{SQD}}=1.0\) and \(P_{\mathrm{feas}}^{\mathrm{SQD}}=1.0\) for every analysed case, despite the direct AE optimal probabilities ranging only from \(0.123\) to \(0.368\). At the same time, the retained subspace becomes proportionally much smaller as \(N\) increases, reaching a \(98.9\%\) dimensional reduction for \(N=12\), while still retaining the optimal state. The additional SQD computational cost is negligible relative to AE, remaining below \(0.02\)~s in all cases.


\subsection{PCE+SQD UC}
\label{subsec:PCE_SQD_Results}

\begin{table*}[h!]
\centering
\scriptsize
\renewcommand{\arraystretch}{1.05}
\resizebox{\textwidth}{!}{%
\begin{tabular}{c c c c c c c c c c c}
\hline
\(N\) &
\(\rho_{\mathrm{eff}}\) &
\(|S|\) &
Red. (\%) &
AE T (s) &
SQD T (s) &
\(P_{\mathrm{opt}}^{\mathrm{AE}}\) &
\(P_{\mathrm{opt}}^{\mathrm{SQD}}\) &
\(P_{\mathrm{feas}}^{\mathrm{AE}}\) &
\(P_{\mathrm{feas}}^{\mathrm{SQD}}\) &
\(C_{\mathrm{feas}}\) \\
\hline
4  & 0.500 & 8  & 50.0 & 15.50  & 0.002 & 0.368 & 1.0 & 0.470 & 1.0 & \(2/3\)   \\
5  & 0.250 & 8  & 75.0 & 25.57  & 0.002 & 0.251 & 1.0 & 0.517 & 1.0 & \(4/4\)   \\
6  & 0.125 & 8  & 87.5 & 31.34  & 0.001 & 0.219 & 1.0 & 0.412 & 1.0 & \(4/5\)   \\
8  & 0.125 & 32 & 87.5 & 54.84  & 0.002 & 0.123 & 1.0 & 0.218 & 1.0 & \(3/9\)   \\
10 & 0.034 & 35 & 96.6 & 80.62  & 0.004 & 0.195 & 1.0 & 0.361 & 1.0 & \(7/11\)  \\
12 & 0.011 & 45 & 98.9 & 268.28 & 0.019 & 0.146 & 1.0 & 0.328 & 1.0 & \(10/13\) \\
\hline
\end{tabular}
}
\vspace{1mm}
\caption{Summary of the selected SQD operating point for the UC case studies. The table reports the effective retained fraction \(\rho_{\mathrm{eff}}\), retained subspace size \(|S|\), dimensional reduction, feasible-state coverage \(C_{\mathrm{feas}}\), feasible and optimal probabilities before and after SQD, and the corresponding AE and SQD computational times. The feasible-state coverage is reported as the number of feasible states retained in the SQD subspace divided by the total number of feasible states. All cases use \(\beta=1.0\), \(\rho_{\mathrm{SQD}}=0.125\), \(T=100\), \(\Delta t=0.1\), \(\eta=0.5\), and \(\tau=1\).}
\label{tab:SQD_summary}
\end{table*}

This subsection evaluates the complete PCE+SQD workflow. PCE first compresses the original \(N\)-qubit Hamiltonian into an \(m\)-qubit representation, after which the adiabatic evolution is performed in the reduced space. SQD is then applied to the sampled reduced distribution. For each system size, \(m\) is fixed to the configuration selected in Section~\ref{subsec:PCE_Results}, while SQD uses \(\beta=1.0\) and \(\rho_{\mathrm{SQD}}=1/8\).

\begin{figure}[!ht]
    \centering
    \includegraphics[width=1\linewidth]{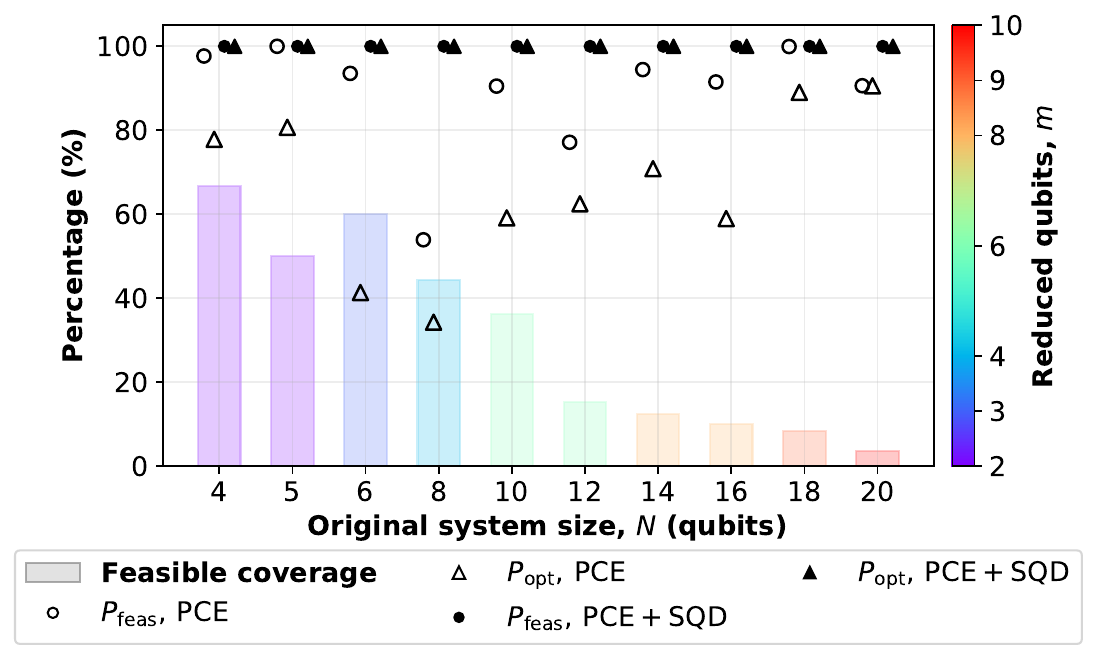}
    \caption{Combined PCE+SQD results for the UC case studies. Bars represent the feasible-state coverage of the retained SQD subspace. Empty markers show the feasible and optimal probabilities after PCE-based adiabatic evolution, while filled markers show the corresponding probabilities after SQD post-processing. The colour scale indicates the reduced number of qubits \(m\).}
    \label{fig:pce_sqd}
\end{figure}

Figure~\ref{fig:pce_sqd} summarises the combined results. The bars indicate the feasible-state coverage of the retained SQD subspace. Empty markers show the feasible and optimal probabilities after PCE-based adiabatic evolution, while filled markers show the corresponding probabilities after SQD post-processing.

The combined workflow consistently refines the PCE distribution, with SQD recovering \(P_{\mathrm{opt}}=1.0\) and \(P_{\mathrm{feas}}=1.0\) for all analysed instances from \(N=4\) to \(N=20\). This is achieved even when the PCE probabilities are moderate or low, showing that the optimal configuration remains represented in the reduced sampled space. At the same time, the retained feasible state coverage decreases with problem size, confirming that full feasible coverage is not required for successful recovery of the optimum. Overall, PCE provides the dimensional reduction required to extend the adiabatic simulations to larger UC instances, while SQD acts as an effective low-cost refinement stage for the compressed solution distribution.

\section{Discussion}

The results obtained across the four analysed workflows highlight the complementary roles of PCE and SQD in the proposed quantum UC framework. Direct AE provides the reference behaviour of the original \(N\)-qubit formulation, PCE reduces the dimension of the Hamiltonian before the adiabatic evolution, and SQD refines the information contained in the resulting sampled distribution. The two techniques therefore address different limitations of the direct formulation.

The PCE results demonstrate that substantial qubit reductions can be achieved without systematically degrading solution quality. The selected configurations reduce the evolved dimension by approximately \(40\%\)-\(60\%\), while \(P_{\mathrm{opt}}\) remains between \(0.343\) and \(0.906\) across the analysed cases. In particular, the larger \(N=18\) and \(N=20\) instances achieve \(P_{\mathrm{opt}}=0.890\) and \(0.906\), respectively, despite a \(50\%\) qubit reduction. The feasible probability also remains high for most cases, although the \(N=8\) instance shows that the performance of the compressed evolution remains instance-dependent.

Feasible-state coverage provides an additional indication of the selectivity introduced by PCE. High probability performance does not require preservation of the complete feasible set: for \(N=18\) and \(N=20\), only \(2/24\) and \(2/28\) feasible schedules are represented, respectively, while the optimal probability remains close to \(0.9\). This confirms that the quality of the reduced representation depends primarily on retaining relevant feasible configurations and on the structure of the resulting reduced Hamiltonian, rather than on maximising feasible-state coverage itself.

\begin{figure*}[!h]
    \centering
    \includegraphics[width=\textwidth]{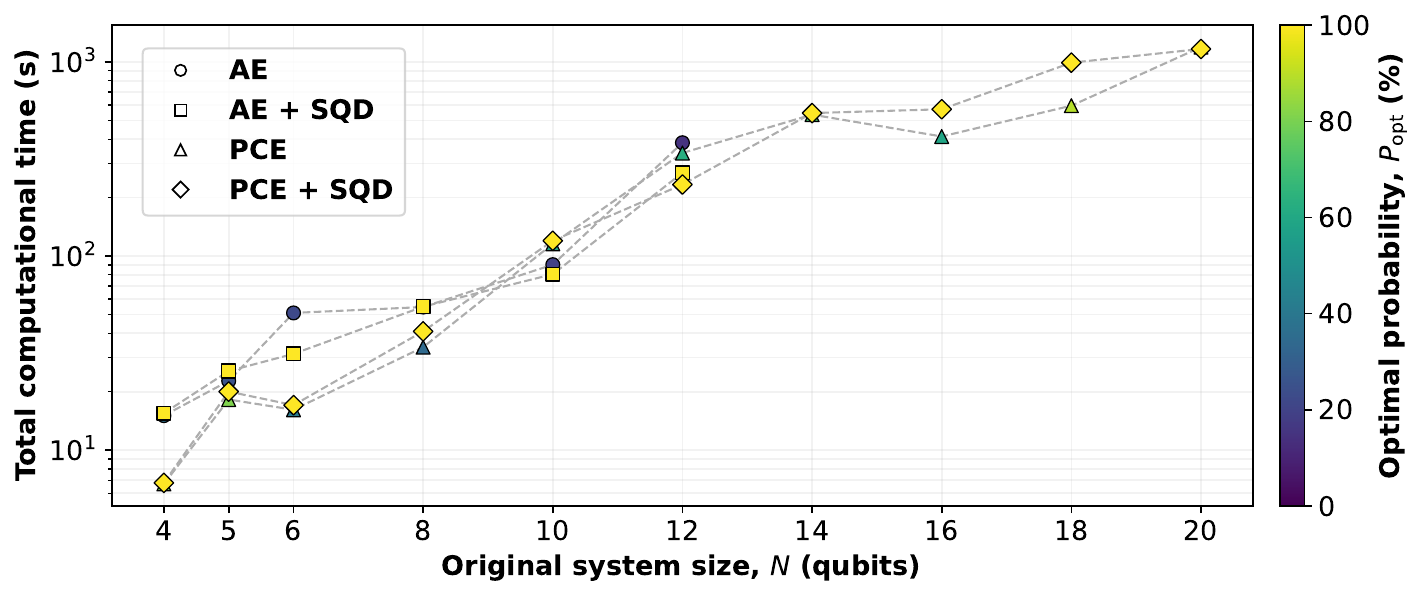}
    \caption{Comparison of total computational time and optimal-solution probability for direct AE, AE+SQD, PCE, and PCE+SQD as the original UC problem size increases.}
    \label{fig:discussion_time_probability}
\end{figure*}

SQD exhibits a different advantage. For the direct AE cases, the resulting distributions contain only moderate feasible and optimal probability, with \(P_{\mathrm{feas}}^{\mathrm{AE}}\) ranging from \(0.218\) to \(0.517\) and \(P_{\mathrm{opt}}^{\mathrm{AE}}\) from \(0.123\) to \(0.368\). After SQD, both probabilities reach \(1.0\) for every analysed instance. This improvement is achieved while retaining only a fraction of the original Hilbert space, with reductions increasing up to \(98.9\%\) for \(N=12\). Therefore, SQD is able to identify and concentrate the relevant solution even when only a reduced subset of the sampled configurations is retained.

The computational overhead introduced by SQD is negligible compared with the adiabatic evolution. Its post-processing time remains below \(0.02\)~s in all direct cases, whereas the AE time increases from approximately \(15\)~s for \(N=4\) to more than \(260\)~s for \(N=12\). SQD should therefore be interpreted as a refinement mechanism rather than a technique for reducing the cost of the preceding quantum evolution.

Combining both techniques preserves these two advantages. PCE first reduces the number of qubits that must be evolved and extends the study to original systems up to \(N=20\), while SQD subsequently concentrates the reduced sampled distribution. Across all analysed system sizes, the combined approach reaches \(P_{\mathrm{opt}}=1.0\) and \(P_{\mathrm{feas}}=1.0\). Consequently, PCE provides the dimensional reduction required for scalability, while SQD improves the reliability of the final extracted solution.

Figure~\ref{fig:discussion_time_probability} provides a joint comparison of computational time and optimal probability for the four workflows.

The runtime results further distinguish the roles of the two techniques. Direct AE and AE+SQD are restricted to \(N\leq12\) in the present emulator because $2^N$-dimensional state vectors surpass memory limits. PCE changes this scaling by performing the adiabatic evolution on an \(m\)-qubit Hamiltonian, enabling the \(N=14\), \(16\), \(18\), and \(20\) cases to be evaluated in the same environment. The measured PCE runtime is not strictly monotonic with \(N\), since it depends on both the selected value of \(m\) and the complexity of the encoded Hamiltonian. SQD adds almost no additional runtime to either the direct or compressed workflow.

\begin{figure}[!ht]
    \centering
    \includegraphics[width=0.85\linewidth]{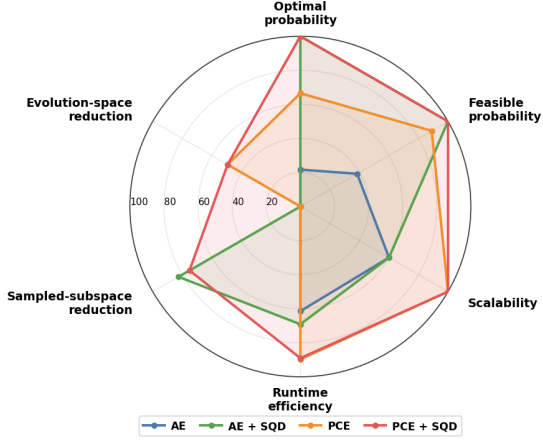}
    \caption{Comparative representation of the main performance indicators for the four analysed UC quantum workflows.}
    \label{fig:discussion_kpi_radar}
\end{figure}

The considered KPIs capture different aspects of the workflow performance. Qubit reduction measures the dimensional benefit provided by PCE, feasible-state coverage characterises the information retained by the reduced representations, sampled-subspace reduction, \(P_{\mathrm{feas}}\) and \(P_{\mathrm{opt}}\) quantify the quality of the resulting distributions, and computational time reflects the practical cost of the simulated workflow. Figure~\ref{fig:discussion_kpi_radar} summarises these complementary behaviours, showing how PCE primarily contributes dimensional reduction and scalability, SQD improves solution concentration, and their combination provides the most balanced performance across the analysed indicators. Considering these indicators jointly, PCE and SQD are therefore better interpreted as complementary stages: PCE addresses dimensional complexity, whereas SQD improves the quality of the solution extracted from the available samples.

For visual comparison, the indicators in Fig.~\ref{fig:discussion_kpi_radar} are normalised to a common 0-100 scale. Runtime efficiency is evaluated over the common cases up to \(N=12\), while scalability reflects the largest original problem size addressed by each workflow relative to \(N=20\).

The reported computational times should nevertheless be interpreted within the limitations of the classical emulator. The restriction of direct AE beyond \(N=12\) results from the exponential memory requirements of state-vector simulation and does not directly represent the behaviour of a physical quantum processor. Likewise, the computational benefit of PCE reflects the reduction of the state space that must be simulated classically. Evaluation on quantum hardware will therefore be required to determine how the observed reductions translate into practical resource and execution-time advantages.

\section{Conclusions and future work}

This work has presented a quantum optimisation workflow for the Unit Commitment problem based on adiabatic quantum evolution and two complementary complexity-reduction techniques: Pauli Correlation Encoding as a pre-processing stage and Sample-based Quantum Diagonalisation as a post-processing stage. Four configurations were analysed: direct AE, AE+SQD, PCE+AE, and the complete PCE+SQD workflow, using UC instances ranging from \(N=4\) to \(N=20\).

The PCE results show that the evolved Hamiltonian can be reduced by approximately \(40\%\)--\(60\%\) while retaining useful feasible and optimal configurations. The largest cases maintain high optimal probabilities after compression, reaching \(P_{\mathrm{opt}}=0.890\) and \(0.906\) for \(N=18\) and \(N=20\), respectively. The results also confirm that high feasible-state coverage is not a prerequisite for good performance, since the most relevant configurations can be preserved even when only a small fraction of the feasible space is encoded.

SQD provides a complementary improvement. In the direct cases, the AE distributions exhibit moderate feasible and optimal probabilities, while SQD increases both \(P_{\mathrm{feas}}\) and \(P_{\mathrm{opt}}\) to \(1.0\) for every analysed instance. This refinement is obtained using strongly reduced sampled subspaces and introduces a negligible computational overhead relative to the adiabatic evolution.

The combined PCE+SQD workflow consequently provides the most complete behaviour among the evaluated configurations. PCE reduces the number of qubits required during the evolution and permits original UC systems up to \(N=20\) to be treated in the present simulation environment, while SQD consistently recovers a fully concentrated feasible and optimal solution from the compressed samples. The two techniques therefore address different aspects of scalability and solution quality and should be regarded as complementary rather than competing approaches.

Future work will investigate the sensitivity and scalability of the proposed workflow with respect to the structure of the feasible space, including variations in the number and distribution of feasible commitment schedules for a fixed problem size. This analysis will help determine how feasible-state density influences both the PCE representation and the effectiveness of the subsequent SQD refinement. The selection of the reduced dimension and the spectral characteristics of the encoded Hamiltonians will also be studied in greater detail.

The UC formulation will additionally be extended towards more realistic operational conditions, including multi-period constraints and additional generator characteristics. Finally, implementation on quantum annealing or gate-based quantum hardware will be required to assess whether the dimensional reductions and sample-based refinement observed under classical emulation translate into practical reductions in quantum resources and execution time.


\section{Appendix}
\subsection{Derivation of the automatic penalty baseline}
\label{app:lambda_auto_derivation}

This appendix summarises the penalty construction used in the normalised UC QUBO formulation. For a binary commitment vector $\mathbf{u}$, let
\begin{equation}
\tilde f(\mathbf{u}) = \sum_{g\in\mathcal{G}} \tilde c_g u_g
\end{equation}
be the normalised UC cost, and let
\begin{equation}
\nu(\mathbf{u})=
\left(
\sum_{g\in\mathcal{G}} \tilde P_g u_g-\tilde D
\right)^2
\end{equation}
be the squared normalised demand-balance violation. The unconstrained objective is then
\begin{equation}
H_{\mathrm{QUBO}}(\mathbf{u})=
\tilde f(\mathbf{u})+\lambda \nu(\mathbf{u}).
\end{equation}
Let $\mathbf{u}^{\star}$ denote the minimum-cost feasible schedule, so that $\nu(\mathbf{u}^{\star})=0$. For any infeasible schedule $\mathbf{u}$ with a lower unpenalised cost than $\mathbf{u}^{\star}$, preserving the feasible optimum as the QUBO ground state requires
\begin{equation}
\tilde f(\mathbf{u}^{\star})
<
\tilde f(\mathbf{u})+\lambda \nu(\mathbf{u}).
\end{equation}
Therefore, the limiting penalty associated with an infeasible competitor is
\begin{equation}
\lambda(\mathbf{u})=
\frac{
\tilde f(\mathbf{u}^{\star})-\tilde f(\mathbf{u})
}{
\nu(\mathbf{u})
},
\qquad \nu(\mathbf{u})>0.
\end{equation}
The minimum penalty that removes all lower-cost infeasible competitors is taken as
\begin{equation}
\lambda_{\min}
=
\max_{\mathbf{u}\notin\mathcal{F}}
\left\{
0,
\frac{
\tilde f(\mathbf{u}^{\star})-\tilde f(\mathbf{u})
}{
\nu(\mathbf{u})
}
\right\}.
\end{equation}
To obtain a common automatic scaling across the analysed instances, the baseline used in this work is
\begin{equation}
\lambda_{\mathrm{auto}}
=
2\lambda_{\min}
\left(1+\frac{1}{N}\right),
\end{equation}
where the factor $(1+1/N)$ introduces a dimension-dependent safety margin that is larger for small problems and approaches one as the number of logical qubits increases. The effective penalty is finally defined as
\begin{equation}
\lambda=\eta\lambda_{\mathrm{auto}}.
\end{equation}
Using the common value $\eta=0.5$ adopted in the case studies gives
\begin{equation}
\lambda
=
\lambda_{\min}
\left(1+\frac{1}{N}\right),
\end{equation}
which places the effective penalty above the limiting value $\lambda_{\min}$ while avoiding an excessive increase with problem dimension. This is the same penalty construction used in the numerical experiments reported in the main text.

\EOD

\begin{thebibliography}{10}
\providecommand{\url}[1]{#1}
\csname url@samestyle\endcsname
\providecommand{\newblock}{\relax}
\providecommand{\bibinfo}[2]{#2}
\providecommand{\BIBentrySTDinterwordspacing}{\spaceskip=0pt\relax}
\providecommand{\BIBentryALTinterwordstretchfactor}{4}
\providecommand{\BIBentryALTinterwordspacing}{\spaceskip=\fontdimen2\font plus
\BIBentryALTinterwordstretchfactor\fontdimen3\font minus
  \fontdimen4\font\relax}
\providecommand{\BIBforeignlanguage}[2]{{%
\expandafter\ifx\csname l@#1\endcsname\relax
\typeout{** WARNING: IEEEtran.bst: No hyphenation pattern has been}%
\typeout{** loaded for the language `#1'. Using the pattern for}%
\typeout{** the default language instead.}%
\else
\language=\csname l@#1\endcsname
\fi
#2}}
\providecommand{\BIBdecl}{\relax}
\BIBdecl

\bibitem{ganeshamurthy2024next}
P.~A. Ganeshamurthy, K.~Ghosh, C.~O'Meara, G.~Cortiana, J.~Schiefelbein-Lach,
  and A.~Monti, ``Next generation power system planning and operation with
  quantum computation,'' \emph{IEEE Access}, vol.~12, pp. 182\,673--182\,692,
  2024.

\bibitem{Liu2023}
H.~Liu and W.~Tang, ``Quantum computing for power systems: Tutorial, review,
  challenges, and prospects,'' \emph{Electric Power Systems Research}, vol.
  223, p. 109530, 2023.

\bibitem{Ullah2022}
M.~H. Ullah, R.~Eskandarpour, H.~Zheng, and A.~Khodaei, ``Quantum computing for
  smart grid applications,'' \emph{IET Generation, Transmission \&
  Distribution}, vol.~16, no.~21, pp. 4239--4257, 2022.

\bibitem{amani2023quantum}
F.~Amani, R.~Mahroo, and A.~Kargarian, ``Quantum-enhanced {DC} optimal power
  flow,'' in \emph{2023 IEEE Texas Power and Energy Conference (TPEC)}, 2023,
  pp. 1--6.

\bibitem{Carrillo}
M.~Carrillo-Mu{\~n}oz, S.~Martinez-Hermida, S.~Barja-Martinez, A.~E.
  Salda{\~n}a-Gonz{\'a}lez, and M.~Arag{\"u}{\'e}s~Pe{\~n}alba, ``Quantum
  optimization approach based on {Qibo} framework for {DC-OPF},'' in \emph{2025
  IEEE PES Innovative Smart Grid Technologies Conference Europe (ISGT Europe)},
  2025, pp. 1--5.

\bibitem{liu2024quantum}
J.~Liu, H.~Zheng, M.~Hanada, K.~Setia, and D.~Wu, ``Quantum power flows: From
  theory to practice,'' \emph{Quantum Machine Intelligence}, vol.~6, no.~2,
  p.~55, 2024.

\bibitem{bucher2025grid}
D.~Bucher, D.~Porawski, B.~Wimmer, J.~N{\"u}{\ss}lein, C.~O'Meara, G.~Cortiana,
  and C.~Linnhoff-Popien, ``Grid cost allocation in peer-to-peer electricity
  markets: Benchmarking classical and quantum optimization approaches,'' in
  \emph{Proceedings of the 17th International Conference on Agents and
  Artificial Intelligence -- Volume 1: QAIO}.\hskip 1em plus 0.5em minus
  0.4em\relax SciTePress, 2025, pp. 751--762.

\bibitem{o2023quantum}
C.~O'Meara, M.~Fern{\'a}ndez-Campoamor, G.~Cortiana, and J.~Bernab{\'e}-Moreno,
  ``Quantum software architecture blueprints for the cloud: Overview and
  application to peer-2-peer energy trading,'' in \emph{2023 IEEE Conference on
  Technologies for Sustainability (SusTech)}, 2023, pp. 191--198.

\bibitem{nikmehr2022quantum}
N.~Nikmehr, P.~Zhang, and M.~A. Bragin, ``Quantum distributed unit commitment:
  An application in microgrids,'' \emph{IEEE Transactions on Power Systems},
  vol.~37, no.~5, pp. 3592--3603, 2022.

\bibitem{koretsky2021adapting}
S.~Koretsky, P.~Gokhale, J.~M. Baker, J.~Viszlai, H.~Zheng, N.~Gurung, R.~Burg,
  E.~A. Paaso, A.~Khodaei, R.~Eskandarpour, and F.~T. Chong, ``Adapting quantum
  approximation optimization algorithm ({QAOA}) for unit commitment,'' in
  \emph{2021 IEEE International Conference on Quantum Computing and Engineering
  (QCE)}, 2021, pp. 181--187.

\bibitem{ling2025hybrid}
J.~Ling, Q.~Zhang, G.~Geng, and Q.~Jiang, ``Hybrid quantum annealing
  decomposition framework for unit commitment,'' \emph{Electric Power Systems
  Research}, vol. 238, p. 111121, 2025.

\bibitem{hong2025qubit}
W.~Hong, W.~Xu, and F.~Teng, ``Qubit-efficient quantum annealing for stochastic
  unit commitment,'' 2025.

\bibitem{soloviev2025large}
V.~P. Soloviev and M.~Krompiec, ``Large-scale portfolio optimization using
  pauli correlation encoding,'' 2025.

\bibitem{padin2026pauli}
J.~Pad{\'i}n-Mart{\'i}nez, V.~P. Soloviev, A.~Borrallo-Rentero,
  A.~Rodr{\'i}guez-Otero, R.~Alfonso-Rodr{\'i}guez, and M.~Krompiec, ``Pauli
  correlation encoding for budget-constrained optimization,'' 2026.

\bibitem{wang2025sample}
Q.~Wang, M.~Motta, R.~D'Cunha, K.~J. Sung, M.~R. Hermes, T.~Gujarati,
  Y.~Kawashima, Y.-y. Ohnishi, G.~O. Jones, and L.~Gagliardi, ``Sample-based
  quantum diagonalization as parallel fragment solver for the localized active
  space self-consistent field method,'' 2025.

\bibitem{sciorilli2025towards}
M.~Sciorilli, L.~Borges, T.~L. Patti, D.~Garc{\'i}a-Mart{\'i}n, G.~Camilo,
  A.~Anandkumar, and L.~Aolita, ``Towards large-scale quantum optimization
  solvers with few qubits,'' \emph{Nature Communications}, vol.~16, p. 476,
  2025.

\bibitem{robledo2025sqd}
J.~Robledo-Moreno, M.~Motta, H.~Haas, A.~Javadi-Abhari, P.~Jurcevic, W.~Kirby,
  S.~Martiel, K.~Sharma, S.~Sharma, T.~Shirakawa, I.~Sitdikov, R.-Y. Sun, K.~J.
  Sung, M.~Takita, M.~C. Tran, S.~Yunoki, and A.~Mezzacapo, ``Chemistry beyond
  the scale of exact diagonalization on a quantum-centric supercomputer,''
  \emph{Science Advances}, vol.~11, no.~25, p. eadu9991, 2025.

\bibitem{kaliakin2025implicit_sqd}
D.~S. Kaliakin, A.~Shajan, F.~Liang, and K.~M.~J. Merz, ``Implicit solvent
  sample-based quantum diagonalization,'' \emph{The Journal of Physical
  Chemistry B}, 2025.

\bibitem{shajan2025sqd_dmet}
A.~Shajan, D.~S. Kaliakin, A.~Mitra, J.~Robledo-Moreno, Z.~Li, M.~Motta,
  C.~Johnson, A.~A. Saki, S.~Das, I.~Sitdikov, A.~Mezzacapo, and K.~M.~J. Merz,
  ``Toward quantum-centric simulations of extended molecules: Sample-based
  quantum diagonalization enhanced with density matrix embedding theory,''
  \emph{Journal of Chemical Theory and Computation}, vol.~21, no.~14, pp.
  6801--6810, 2025.

\bibitem{tan2021qubit_efficient_encoding}
B.~Tan, M.-A. Lemonde, S.~Thanasilp, J.~Tangpanitanon, and D.~G. Angelakis,
  ``Qubit-efficient encoding schemes for binary optimisation problems,''
  \emph{Quantum}, vol.~5, p. 454, 2021.

\bibitem{fuller2024quantum}
B.~Fuller, C.~Hadfield, J.~R. Glick, T.~Imamichi, T.~Itoko, R.~J. Thompson,
  Y.~Jiao, M.~M. Kagele, A.~W. Blom-Schieber, R.~Raymond, and A.~Mezzacapo,
  ``Approximate solutions of combinatorial problems via quantum relaxations,''
  \emph{IEEE Transactions on Quantum Engineering}, 2024.

\bibitem{kondo2025recursive}
R.~Kondo, Y.~Sato, R.~Raymond, and N.~Yamamoto, ``Recursive quantum relaxation
  for combinatorial optimization problems,'' \emph{Quantum}, vol.~9, p. 1594,
  2025.

\bibitem{egger2021warm_starting}
D.~J. Egger, J.~Mare{\v{c}}ek, and S.~Woerner, ``Warm-starting quantum
  optimization,'' \emph{Quantum}, vol.~5, p. 479, 2021.

\bibitem{muller2026quantum_annealing_unit_scheduling}
S.~M{\"u}ller, M.~Dukalski, and F.~Phillipson, ``Quantum annealing for
  optimizing unit scheduling in renewable energy systems: Formulation and
  evaluation,'' \emph{IEEE Transactions on Power Systems}, vol.~41, no.~2, pp.
  836--846, 2026.

\bibitem{feng2023quantum_surrogate_uc}
F.~Feng, P.~Zhang, M.~A. Bragin, and Y.~Zhou, ``Novel resolution of unit
  commitment problems through quantum surrogate lagrangian relaxation,''
  \emph{IEEE Transactions on Power Systems}, vol.~38, no.~3, pp. 2460--2471,
  2023.

\bibitem{mahroo2022hybrid_quantum_uc}
R.~Mahroo and A.~Kargarian, ``Hybrid quantum-classical unit commitment,'' in
  \emph{2022 IEEE Texas Power and Energy Conference (TPEC)}, 2022, pp. 1--5.

\bibitem{paterakis2023hybrid_benders_uc}
N.~G. Paterakis, ``Hybrid quantum-classical multi-cut benders approach with a
  power system application,'' \emph{Computers \& Chemical Engineering}, vol.
  172, p. 108161, 2023.

\bibitem{koch2025hubo}
F.~Koch, S.~Panahiyan, R.~Mukherjee, J.~Doetsch, and D.~Jaksch,
  ``Resource-efficient quantum optimization via higher-order encoding,'' 2025.

\bibitem{sciorilli2025competitive}
M.~Sciorilli, G.~Camilo, T.~O. Maciel, A.~Canabarro, L.~Borges, and L.~Aolita,
  ``A competitive {NISQ} and qubit-efficient solver for the {LABS} problem,''
  2025.

\bibitem{sharma2025comparative}
M.~Sharma and H.~C. Lau, ``A comparative study of quantum optimization
  techniques for solving combinatorial optimization benchmark problems,'' 2025.

\bibitem{nakaji2022approximate}
K.~Nakaji, S.~Uno, Y.~Suzuki, R.~Raymond, T.~Onodera, T.~Tanaka, H.~Tezuka,
  N.~Mitsuda, and N.~Yamamoto, ``Approximate amplitude encoding in shallow
  parameterized quantum circuits and its application to financial market
  indicators,'' \emph{Physical Review Research}, vol.~4, no.~2, p. 023136,
  2022.

\bibitem{zhou2022quantum_power_systems}
Y.~Zhou, Z.~Tang, N.~Nikmehr, P.~Babahajiani, F.~Feng, T.-C. Wei, H.~Zheng, and
  P.~Zhang, ``Quantum computing in power systems,'' \emph{iEnergy}, vol.~1,
  no.~2, pp. 170--187, 2022.

\bibitem{morstyn2024quantum_net_zero}
T.~Morstyn and X.~Wang, ``Opportunities for quantum computing within net-zero
  power system optimization,'' \emph{Joule}, vol.~8, no.~6, pp. 1619--1640,
  2024.

\bibitem{qibo_paper}
S.~Efthymiou, S.~Ramos-Calderer, C.~Bravo-Prieto, A.~P{\'e}rez-Salinas,
  D.~Garc{\'i}a-Mart{\'i}n, A.~Garcia-Saez, J.~I. Latorre, and S.~Carrazza,
  ``Qibo: A framework for quantum simulation with hardware acceleration,''
  \emph{Quantum Science and Technology}, vol.~7, no.~1, p. 015018, 2021.

\bibitem{qibojit_paper}
S.~Efthymiou, M.~Lazzarin, A.~Pasquale, and S.~Carrazza, ``Quantum simulation
  with just-in-time compilation,'' \emph{Quantum}, vol.~6, p. 814, 2022.

\bibitem{qibo}
Qibo, ``Qibo,'' [Online]. Available: \url{https://qibo.science/}.

\end{thebibliography}
\end{document}